\documentclass[
preprint,
amsmath,amssymb,
aps,
]{revtex4-2}

\usepackage{graphicx}
\usepackage{dcolumn}
\usepackage{bm}
\usepackage{hyperref}
\usepackage[dvipsnames]{xcolor}
\usepackage{comment}
\usepackage{multirow}
\usepackage{mathtools}
\usepackage{xcolor}
\usepackage{bigstrut}
\usepackage{booktabs}
\usepackage{array}

\usepackage{slashed}
\usepackage{mathrsfs}

\newcommand{\be}{\begin{equation}}
\newcommand{\ee}{\end{equation}}
\newcommand{\bea}{\begin{eqnarray}}
\newcommand{\eea}{\end{eqnarray}}
\newcommand{\beq}{\begin{equation}}
\newcommand{\eeq}{\end{equation}}
\def\bsp#1\esp{\begin{split}#1\end{split}}
\def\bal#1\eal{\begin{align}#1\end{align}}

\newcommand\bom[1]     {{\mbox{\boldmath $#1$}}}

\newcommand\tS   {\theta_S}
\newcommand{\tG}{\theta_G}

\newcommand\rL   {\ensuremath{\mathrm{L}}}
\newcommand\rR   {\ensuremath{\mathrm{R}}}

\newcommand\MeV  {\ensuremath{\mathrm{MeV}}}
\newcommand\GeV  {\ensuremath{\mathrm{GeV}}}
\newcommand\TeV  {\ensuremath{\mathrm{TeV}}}
\newcommand\rc   {\ensuremath{\mathrm{c}}}

\newcommand\ri   {\ensuremath{\mathrm{i}}}

\newcommand\rs   {\ensuremath{\mathrm{s}}}

\newcommand\gL   {\ensuremath{g_\mathrm{L}}}

\newcommand\gz   {g_{z}}

\newcommand\gZ   {\ensuremath{g_{Z^0}}}

\newcommand\cL   {\ensuremath{\mathcal{L}}}
\newcommand{\nub}{\nu}
\newcommand\Mpl {\ensuremath{{M}_{\rm Pl}}}

\newcommand\cw {\ensuremath{ \rc_W}}
\newcommand\sw {\ensuremath{ \rs_W}}
\newcommand\cz {\ensuremath{ \rc_Z}}
\newcommand\sz {\ensuremath{ \rs_Z}}

\newcommand\cs {\ensuremath{ \rc_S}}
\newcommand\sss {\ensuremath{ \rs_S}}
\newcommand\tanb {\ensuremath{ \tan\beta}}

\makeatletter
\newcommand{\undersim}[1]{\mathrel{\mathpalette\@undersim{#1}}}
\newcommand{\@undersim}[2]{%
  \vcenter{%
    \ialign{%
      ##\cr
      $\m@th#1#2$\cr
      \noalign{\nointerlineskip\kern.2ex}
      $\m@th#1\sim$\cr
      \noalign{\kern-.4ex}
    }%
  }%
}

\makeatother

\begin{document}

\title{Exclusion bounds for neutral gauge bosons}

\author{Zolt\'an P\'eli}
\email{zoltan.peli@ttk.elte.hu}
\affiliation{Institute for Theoretical Physics, ELTE E\"otv\"os Lor\'and University,
P\'azm\'any P\'eter s\'et\'any 1/A, 1117 Budapest, Hungary
}
\author{Zolt\'an Tr\'ocs\'anyi}
\email{zoltan.trocsanyi@cern.ch}
\affiliation{Institute for Theoretical Physics, ELTE E\"otv\"os Lor\'and University and
HUN-REN ELTE Theoretical Physics Research Group,
P\'azm\'any P\'eter s\'et\'any 1/A, 1117 Budapest, Hungary, also at\\
University of Debrecen, Bem t\'er 18/A, 4026 Debrecen, Hungary
}
\date{\today}
\begin{abstract}
We study how the recent experimental results constrain the 
gauge sectors of U(1) extensions of the standard model
using a novel representation of the parameter space.
We determine the bounds on the mixing angle between the massive 
gauge bosons, or equivalently, the new gauge coupling as a function of 
the mass $M_{Z'}$ of the new neutral gauge boson $Z'$ in the 
approximate range $(10^{-2},10^4)$\,GeV/$c^2$. 
We consider the most stringent bounds obtained from direct 
searches for the $Z'$. We also exhibit the allowed parameter 
space by comparing the predicted and measured values of the 
$\rho$ parameter and those of the mass of the $W$ boson. 
Finally, we discuss the prospects of $Z'$ searches at future colliders.
\end{abstract}

\maketitle

\vspace*{-2em}
\section{Introduction}

The standard model of particle interactions (SM) has been tested to high precision 
both in low energy experiments as well as at high energy colliders 
\cite{Workman:2022ynf}. Most recently the large LHC experiments find spectacular 
agreement between their experimental results and the SM predictions 
\cite{ATLAS:2022djm,CMS-SM}, which leaves very little room for new physics.
Nevertheless, we do not have doubt that the SM cannot describe all observations
in the microworld. Most notably, the masses of neutrinos, the baryon asymmetry
and the origin of dark matter in the Universe are clear indications of the need
for physics beyond the standard model (BSM). The nature of this new physics however, 
remains elusive.

There is also a 5$\sigma$ tension between the measured value of the muon 
anomalous magnetic moment $a_\mu$ \cite{Muong-2:2023cdq} and the SM prediction 
when the hadronic vacuum polarization to the photon is extracted from 
the measured total hadronic cross section in electron-positron annihilation
at low energies \cite{Aoyama:2020ynm}. A natural explanation for such
a difference is the contribution of a new heavy neutral gauge boson $Z'$
\cite{Pospelov:2008zw,Davoudiasl:2012ig,Cadeddu:2021dqx}. Presently however,
the size of the deviation between the SM prediction and the measurement
is heavily debated. The Budapest-Marseille-Wuppertal lattice collaboration
computed the hadronic vacuum polarization of the photon from first principles
\cite{Borsanyi:2020mff}, and found a much less significant ($<2\sigma$)
tension for $a_\mu$ between theory and experiment. Nevertheless, the exploration
of the effects of a $Z'$ on measurements is interesting because $U(1)_z$ 
extensions provide the simplest possible way to explain a potential fifth 
fundamental force. 

Due to their simplicity, $U(1)_z$ extensions have a more then forty year old 
history \cite{Okun:1982xi} and remain popular \cite{Workman:2022ynf} at present. 
They have been investigated since the operation of the experiments of the Large Electron Positron 
collider in various forms, like gauged extra U(1) symmetry \cite{He:1990pn}, 
which can also be broken by a new scalar \cite{Gopalakrishna:2008dv}, giving rise 
to a new massive gauge boson, often call dark photon \footnote{After 
diagonalization of the mass matrix of the neutral gauge bosons}, $A'$. Lately, 
more complete U(1) extensions of the SM have been studied with the goal of 
explaining several beyond the standard model phenomena simultaneously, 
such as the superweak extension of the standard model (SWSM) \cite{Trocsanyi:2018bkm}.
Even the complete one-loop renormalization of the Dark Abelian Sector 
Model with identical gauge and scalar sectors as in the SWSM, but with 
somewhat different fermion content has been carried out in Ref.~\cite{Dittmaier:2023ovi}.

The continuing theoretical interest is met with similarly ubiquitous experimental
searches for dark photons or more generally, new neutral gauge bosons.
The experimental limits are typically presented in the parameter space of the 
dark photons, which provide serendipitous discovery potential for other
types of vector particles \cite{Ilten:2018crw}.
Constraints have been placed on visible $A'$ decays by 
beam-dump, 
collider, 
fixed-target,
and rare-meson-decay experiments, 
as well as on invisible $A'$  decays. 
New experiments have also been proposed to explore further the parameter
space in the future \cite{FCC:2018bvk,Feng:2017uoz}.

In this work we discuss the presently most constraining experimental 
limits in some part of the parameter space of general U(1) extensions 
that contain right-handed neutrinos, not charged under the SM 
interactions, in the particle spectrum. We focus on
two regions: the case of light and heavy $Z'$, $\xi = M_{Z'}/M_Z \ll 1$ 
and $\xi \gg 1$, or quantitatively $M_{Z'} \in [0.02,10]$\,GeV and 
$M_{Z'} \in [0.2,5]$\,TeV \footnote{The cases of an intermediate mass $Z'$, 
$M_{Z'} \approx M_Z$, and a very light $Z'$, $M_{Z'} \lesssim 20$\,MeV have 
been discussed elsewhere.}. 
The most stringent limits  for a heavy $Z'$
are provided by direct searches at the LHC in Drell-Yan 
pair production $pp\to Z'+X \to \ell^+ \ell^- + X$ \cite{ATLAS:2019erb,CMS:2021ctt}. 
For a light $Z'$ those have been obtained in 
direct searches for an invisibly decaying dark photon in the NA64 
\cite{NA64:2019imj} and BaBar \cite{BaBar:2017tiz} experiments, 
as well as for a dark photon decaying into an electron-positron pair 
in the FASER detector \cite{FASER:2023tle}. Similar studies have 
already been published. For instance, Refs.~\cite{Asai:2022zxw,Asai:2023xxl} 
focus on $U(1)_z$ extensions with a selection of benchmark $z$ charges 
in the light $M_{Z'}$ region. In the present work we use a 
parametrization of the $z$ charges valid for any charge assignments 
that satisfy the conditions of anomaly cancellations and gauge 
invariance. The exclusion limits depend on a 
specific single combination of the free $z$ charges such that 
we also take into account the uncertainty due to the choice of the
renormalization scale where the $z$ charges are set, neglected in Refs.~\cite{Asai:2022zxw,Asai:2023xxl}

As one might expect, in the regions far away from the mass of the $Z$ boson
the mixing angle $\theta_Z$ between the massive neutral gauge bosons is 
small experimentally (below $10^{-3}$), so that one can use expansions 
around $\theta_Z = 0$. In this limit the couplings of the $Z'$ to chiral 
fermions are approximately vector like and universal in the sense that
they all depend on a unique combination of the $z$ 
charge of the right-handed neutrinos and Brout-Englert-Higgs (BEH) 
field. We also discuss how results of electroweak precision 
measurements constrain the parameter space.

\section{Model definition}
\label{sect:u1_extension}

We consider the extensions of the standard model by a $U(1)_z$ 
gauge group with a complex scalar field $\chi$ and three generations 
of right handed neutrinos. The new fields are neutral under the 
standard model gauge interactions. An example for such a model is the 
superweak extension of the standard model (SWSM) 
\cite{Trocsanyi:2018bkm}. However, in the present work we require only 
gauge and gravity anomaly cancellation, otherwise leave the 
$z$ charges arbitrary. In this section we collect the details 
of the model only to the extent used in the present analyses.

\subsection{Scalar sector}
\label{sec:scalars}

In the scalar sector, in addition to the $SU(2)_\rL$-doublet 
Brout-Englert-Higgs field
\begin{equation}
\phi=\left(\!\!\begin{array}{c}
               \phi^{+} \\
               \phi^{0}
             \end{array}\!\!\right) = 
\frac{1}{\sqrt{2}}
\left(\!\!\begin{array}{c}
          \phi_{1}+\ri\phi_{2} \\
          \phi_{3}+\ri\phi_{4}
        \end{array}\!\!
\right)
\,,
\end{equation}
the model contains a complex scalar SM singlet $\chi$.
The Lagrangian of the scalar fields contains the potential energy
\beq
\bsp
V(\phi,\chi) &= - \mu_\phi^2 |\phi|^2 - \mu_\chi^2 |\chi|^2
+ \left(|\phi|^2, |\chi|^2\right)
\left(\!\!\begin{array}{cc}
 \lambda_\phi & \frac{\lambda}2 \\ \frac{\lambda}2 & \lambda_\chi
\end{array}\!\!\right)
\left(\!\!\begin{array}{c}
|\phi|^2 \\ |\chi|^2
\end{array}\!\!\right) 
\\ & \subset  -\cL
\label{eq:V}
\esp
\eeq
where $|\phi|^2 =|\phi^+|^2 + |\phi^0|^2$.
After spontaneous symmetry breaking, we parametrize the scalar fields as
\beq
\bsp
\phi =\frac{1}{\sqrt{2}}\binom{-\ri \sqrt{2}\sigma^+}{v+h'+ \ri\sigma_\phi}
\,,\quad
\chi = \frac{1}{\sqrt{2}}(w + s' + \ri\sigma_\chi)
\esp
\eeq
where $v$ and $w$ are the vacuum expectation values (VEVs) of $\phi$ and $\chi$.
The fields $h'$ and $s'$ are real scalars, $\sigma^+$ is a charged, while 
$\sigma_\phi$ and $\sigma_\chi$ are neutral Goldstone bosons that are gauge eigenstates. 

The gauge and mass eigenstates are related by the rotations
\beq
\binom{h}{s} = \textbf{Z}_S
\binom{h'}{s'}
\,,\quad
\binom{\sigma_Z}{\sigma_{Z'}} = \textbf{Z}_G
\binom{\sigma_\phi}{\sigma_\chi}
\,,
\eeq
with
\beq
\textbf{Z}_X=
\begin{pmatrix}
\cos\theta_X & -\sin\theta_X
\\
\sin\theta_X & ~~ \cos\theta_X
\end{pmatrix}
\eeq
where we denoted the mass eigenstates with $h$, $s$ and $\sigma_Z$,
$\sigma_{Z'}$. The angles $\tS$ and $\tG$ 
are the scalar and Goldstone mixing angles that can
be determined by the diagonalization of the mass matrix of the real scalars 
and that of the neutral Goldstone bosons. In the following, we are going to use 
the abbreviations ${\rm c}_X = \cos\theta_X$ and ${\rm s}_X = \sin\theta_X$ for mixing angles.

\subsection{Gauge sector}

The field strength tensors of the U(1) gauge groups are 
gauge invariant, kinetic mixing is allowed between the gauge fields 
belonging to the hypercharge U(1)$_Y$ and the new U(1)$_z$ gauge 
symmetries. Equivalently, one can choose a basis in which the 
gauge-field strengths do not mix \footnote{This is the convention 
used in \texttt{SARAH} for models with multiple gauged U(1) 
symmetries.}, such that the covariant derivative 
corresponding to the U(1) gauge groups can be parametrized as
\begin{equation}
    D^{\mathrm{U}(1)}_\mu = -\ri \begin{pmatrix}
    y & z
    \end{pmatrix}
    \begin{pmatrix}
    g_{y} &  -g_{yz} \\
    0 &  g_{z}
    \end{pmatrix}
    \begin{pmatrix}
    \cos \alpha &  -\sin\alpha \\
    \sin\alpha &  ~~\cos \alpha 
    \end{pmatrix}
    \begin{pmatrix}
    B_\mu \\ B'_\mu
    \end{pmatrix}
    \label{eq:kineticmixing}
\end{equation}
where $B_\mu$ and $B'_\mu$ are the U(1)$_y$ and U(1)$_z$ gauge fields, 
while  $y$ and $z$ are the corresponding charges. The rotation angle 
$\alpha$ is not physical as it can be absorbed into the definition 
of the gauge fields \cite{Iwamoto:2021fup}. The $y$ charges are the 
eigenvalues of one half times the hypercharge operator $Y$. The $z$ 
charges are assigned such that Yukawa terms including the neutrinos and the 
scalar fields exist, and the gauge and gravity anomalies cancel in each family.
Such a charge assignment can be paramterized with two numbers, 
usually chosen to be the charge of the left-handed quark doublet $z_q$ and 
that of the right-handed u-type quarks $z_u$ \cite{Appelquist:2002mw}.  
The $z$ charge of the field $\chi$ can be fixed without the loss of 
generality as its normalization can be absorbed into the rescaling 
of $g_z$. In this work we use $z_\chi = -1$. 

In general, the $z_i$ charges of the right handed neutrinos have to satisfy 
\be 
\frac{1}{3}\sum_{i=1}^n z_i =  z_u - 4z_q \equiv z_N,
\quad \text{and}\quad
\biggl(\sum_{i=1}^n z_i\biggr)^3 =  9 \sum_{i=1}^n z_i^3.
\ee
A simple and natural choice is to have $n=3$ generations of sterile
neutrinos and generation independent $z$ charges, i.e.~$z_i = z_N$
for any $i=1,2,3$. We find that for phenomenology it is more 
convenient to choose $z_N$ and the $z$ charge of the SM scalar field
$z_\phi$ as independent charges. We exhibit the corresponding $z$ charge
assignment in Table \ref{tab:zchrges}. 
\begin{table}[ht]
\newcommand\zj[1]{$\frac{#1}6$}
\begin{center}
\begin{tabular}{|c||c|c|r|c|}
\hline
\hline
field & $SU(3)_\rc$ & $SU(2)_\rL$ & $y$  & $z$\\
\hline
$Q_\rL$  & 3 & 2 & $\frac16$ &    $z_q = \frac13(z_\phi - z_N)$\\[2mm]
\hline
$U_\rR$  & 3 & 1 & $\frac23$ &    $z_u = \frac13(4z_\phi - z_N)$\\[2mm]
\hline
$D_\rR$  & 3 & 1 &$-\frac13$ & $ z_d =-\frac13(2z_\phi + z_N)$\\[2mm]
\hline
$\ell_\rL$& 1 & 2 &$-\frac12$ & $ z_\ell = z_N - z_\phi$ \\[2mm]
\hline
$N_\rR$  & 1 & 1 & 0  & $z_N$\\[2mm]
\hline
$e_\rR$  & 1 & 1 & $-1$  & $z_e = z_N - 2 z_\phi$ \\[2mm]
\hline
$\phi$  & 1 & 2 & $\frac12$ & $z_\phi$ \\[2mm]
\hline
$\chi$  & 1 & 1 & 0  & $z_\chi = -1$ \\
\hline
\hline
\end{tabular}
\normalsize
\caption{\label{tab:zchrges} Field content and charge assignment of a generic U(1)$_z$ extension of the SM. The field $\phi$ is the Higgs doublet and $\chi$ is a complex scalar field, the rest are Weyl fermions. We show the representations for SU(3)$_\mathrm{c}\otimes$SU(2)$_\mathrm{L}$ and the charges $y$ and $z$ for U(1)$_Y\otimes$U(1)$_z$.
} 
\end{center}
\end{table}

A $D=4$ operator corresponding to a Majorana mass term for the 
sterile neutrinos is allowed only for $z_\chi + 2 z_N = 0$, which 
implies $z_N = 1/2$ with our normalization $z_\chi = -1$. 
For example, in the $B-L$ U(1) extension, $z_N=1/2$ and 
$z_\phi = 0$, while in the SWSM $z_N=1/2$ and $z_\phi = 1$. 

The neutral gauge fields are related to their mass eigenstates 
$A_\mu$, $Z_\mu$ and $Z^{\prime}_\mu$ via two rotations 
\footnote{Our sign convention for $\theta_Z$ agrees with the convention of Ref.~\cite{Trocsanyi:2018bkm}, which differs by a factor of $(-1)$ from that of Ref.~\cite{Iwamoto:2021fup}.}
\beq
\left(\begin{array}{c}
    B_\mu \\
    W^{3}_\mu \\
    B'_\mu
\end{array}\right) =
\left( \begin{array}{ccc}
\cw & -\sw & 0 \\
\sw &  \cw & 0 \\
  0 &  0   &  1 \end{array}
\right)
\left( \begin{array}{ccc}
  1 &   0 &  0   \\
  0 & \cz & -\sz \\
  0 & \sz &  \cz \end{array}
\right)
\left(\begin{array}{c}
    A_\mu \\
    Z_\mu \\
    Z^{\prime}_\mu
\end{array}\right).
\label{eq:gauge-rotation}
\eeq
The two mixing angles are (i) the weak mixing angle $\theta_{\rm W}$ 
and (ii)  the $Z-Z'$ mixing angle $\theta_Z \in[-\pi/4,\pi/4]$.
The former is defined as $\sw = \frac{g_y}{\gZ}$\,, with 
$\gZ^2 = g_y^2+\gL^2$\,, so $e = \gL\sw$ where $\gL$ is the 
SU(2) gauge coupling and $e$ is the elementary charge. The new 
mixing angle is defined as 
\be
\tan(2 \theta_Z) = -\frac{2 \kappa}{1 - \kappa^2 - \tau^2}
\label{eq:tZ}
\ee
in terms of effective couplings
\be
 \kappa = 2 \frac{\gz}{\gZ} z_\phi(\mu)
\text{~~and~~}
\tau = 2\frac{\gz}{\gZ}\tanb 
 \label{eq:kappa-tau}
\ee
where $\tanb = \frac{w}{v}$, and we introduced the effective charge
\be
z_\phi(\mu) = z_\phi - \frac{g_{yz}}{2\gz}
\label{eq:zphi(mu)}
\ee
as the charge $z_\phi$ appears always together with this ratio of the new 
couplings throughout our computations. In Eq.~\eqref{eq:zphi(mu)} we 
indicated the dependence of the couplings on the renormalization scale 
$\mu$ to emphasize the scale dependence of the effective charge
defined as abbreviation. 
It is possible to choose a basis of the fundamental gauge fields such 
that $g_{yz}(\mu_0) =0$ at a fixed, but arbitrary renormalization 
scale $\mu_0$. Clearly, $z_\phi(\mu_0) = z_\phi$ at this scale, 
i.e.~$\mu_0$ is the scale where all $z$ charges are set. The scale
$\mu_0$ can be chosen at will, but the running of $z_\phi(\mu)$ 
introduces some theoretical uncertainty to our predictions, 
whose size depends on the actual choice of $\mu_0$. In order to 
assess this uncertainty, we discuss the one-loop 
running of the ratio $\eta = g_{yz}/\gz$ in
App.~\ref{app:etarunning}.

In terms of the mixing angles and effective couplings, the 
masses of the gauge bosons are $ M_W  = \frac{1}{2}g_{\rm L} v$,
\be
M_Z = \frac{M_W}{\cw} \sqrt{R(\cz,\sz)}
\,,\quad
M_{Z'} = \frac{M_W}{\cw} \sqrt{R(\sz,-\cz)}
\,,
\label{eq:vectormasses}
\ee
with
$R(x,y) = \bigl(x - \kappa y\bigr)^2 + \bigl(\tau y\bigr)^2.$
The coupling parameters $\kappa$ and $\tau$ can be expressed in 
terms of  the experimentally more accessible parameters $M_{Z'}$ 
and $\theta_Z$ as 
\begin{equation}\label{eq:mzpthetaz}
    \kappa = -\cz \sz ~\frac{M_Z^2 - M_{Z'}^2}{ \cz^2 M_Z^2 + \sz^2 M_{Z'}^2}
    ~~~~\text{and}~~~
    \tau = \frac{M_Z M_{Z'}}{ \cz^2 M_Z^2 + \sz^2 M_{Z'}^2}
    \,.
\end{equation}
Taking the ratio $\kappa/\tau$, Eqs.~\eqref{eq:kappa-tau} and \eqref{eq:mzpthetaz} imply  
\be 
-\sz \cz (1- \xi^2)
 = \frac{\xi z_\phi(\mu)}{\tanb}
\label{eq:kappa/tau}
\ee
where we remind the reader that $\xi = M_{Z'}/M_Z$.

\subsection{Modified $\rho$ parameter}
\label{subsect:mod-rho}
The well known SM tree-level relationship between 
the masses of the $W$ and $Z$ bosons is usually expressed as $\rho = 1$ where
\be 
\rho = \frac{M_W^2}{\cw^2 M_Z^2}
\,.
\ee
In the extended model it is no longer equal to one at the tree level as it is modified to
\be 
\label{eq:u1_rho}
\rho = 1 - \sz^2\,(1-\xi^2)
\,.
\ee
Experimentally, from global fits \cite{Workman:2022ynf} one has
\be
\label{eq:u1_rho_exp}
\rho = 1.00038 \pm 0.00020\,,
\ee
which implies that $\sz^2 \ll 1$ for either a light or a heavy $Z'$ boson. 
Utilizing the smallness of $\sz$,
we can also express the $\rho$ parameter in terms of the effective couplings,
$\rho = 1 - \kappa^2/(1-\tau^2) + \mathcal{O}\bigl(\sz^4\bigr)$,
or using the Lagrangian couplings and $M_{Z'}$ as
\be 
\rho =  1 - \frac{v^2}{M_Z^2 -M_{Z'}^2}\big(z_\phi(\mu)\,g_z\big)^2
+ \mathcal{O}\bigl(\sz^4\bigr)
\,,
\ee
which we use below. Equivalently, we can express $\rho$ using $\tanb$. In the 
limit of a heavy $Z'$ we have
$\rho  \simeq 1 + \big(z_\phi(\mu)/\tanb\big)^2$,
whereas a light $Z'$ implies
$\rho  \simeq 1 - \big(z_\phi(\mu)\,\xi/\tanb\big)^2$.

\subsection{Vector-axial vector couplings of the Z-prime boson}
\label{sect:VA-cps}

Direct $Z'$ searches at colliders are most often based on the 
Drell-Yan process, hence on decays of the $Z'$ into fermion 
pairs. The relevant theoretical predictions, discussed in detail 
in the following sections, rely on the interaction of the $Z'$ boson
and the fermions, which in the Dirac basis reads as 
(neglecting the mixing among the neutrinos)
\begin{equation}
\cL_{\text{NC}}^{(Z')} = -\frac{e}{2\sw \cw} Z'_\mu 
\sum_f \bar{f} \gamma^\mu \big(v_{Z',f} - a_{Z',f} \gamma_5 \big) f
\,.
\end{equation}
We recall the vector and axial vector couplings $v_{Z',f}$ 
and $a_{Z',f}$ using a parametrization convenient to our 
analysis in Table \ref{tab:sw-cps}, obtained using the 
chiral couplings presented in App.~\ref{app:Chiralcouplings}.
\begin{table}[ht]
\centering
\begin{tabular}{ |c||c|c| } 
 \hline
 ~~~$f$~~~ & $v_{Z',f}$ & $a_{Z',f}$ \\ 
\hline\hline
$\nu$ &  $-\frac{1}{2}\sz +\frac{1}{2}\Big( -\kappa + 2\frac{\tau}{\tanb}z_N\Big)\cz$ 
& $-\frac{1}{2}\bigl(\sz + \kappa \cz \bigr)$ 
\\ [1ex] \hline
$\ell$ &  $-\Big(-\frac{1}{2}+2\sw^2\Big)\sz + \frac{1}{2}\Big( -3\kappa + 2\frac{\tau}{\tanb} z_N\Big)\cz$ 
& $\frac{1}{2}\bigl(\sz + \kappa \cz \bigr)$ 
\\ [1ex] \hline
$u$ & $-\Big(\frac{1}{2}-\frac{4}{3}\sw^2\Big)\sz + \frac{1}{6}\Big( 5\kappa - 2\frac{\tau}{\tanb}z_N \Big)\cz$ 
& $-\frac{1}{2}\bigl(\sz + \kappa \cz \bigr)$ 
\\ [1ex] \hline
$d$ & $-\Big(-\frac{1}{2}+\frac{2}{3}\sw^2\Big)\sz - \frac{1}{6}\Big( \kappa + 2\frac{\tau}{\tanb}z_N \Big)\cz$ 
& $\frac{1}{2}\bigl(\sz + \kappa \cz \bigr)$ 
\\ [1ex] \hline
\end{tabular}
\caption{The vector and axial-vector couplings of the $Z'$ boson to fermions in U(1)$_z$ extensions of the SM. The corresponding couplings of the $Z$ boson $v_{Z,f}$ and $a_{Z,f}$ are obtained by the replacement $(\cz,\:\sz) \to (\sz,\: -\cz)$ in $v_{Z',f}$ and $a_{Z',f}$.}
\label{tab:sw-cps}
\end{table}
Expanding these couplings in terms of the small parameter $\sz$ one obtains the following expressions (recall that $\xi = M_{Z'}/M_Z$):
\be 
\bsp
v_{Z',\nu} \simeq \frac{z_N\,\xi}{\tanb}
-\frac{1}{2}\sz \xi^2 , &\quad
a_{Z',\nu} \simeq  -\frac{1}{2}\sz \xi^2,
\\
v_{Z',\ell} \simeq \frac{z_N\,\xi}{\tanb}
+\frac{1}{2}\sz \bigl(4\cw^2-3\xi^2\bigr) , &\quad
a_{Z',\ell} \simeq \frac{1}{2}\sz \xi^2,
\\
v_{Z',u} \simeq -\frac{z_N\,\xi}{3\tanb}
+ \frac{1}{6}\sz \bigl(-8\cw^2+5\xi^2\bigr) , &\quad
a_{Z',u} \simeq  -\frac{1}{2}\sz \xi^2,
\\
v_{Z',d} \simeq -\frac{z_N\,\xi}{3\tanb}
+ \frac{1}{6}\sz \bigl(4\cw^2-\xi^2 \bigr), &\quad
a_{Z',d} \simeq   \frac{1}{2}\sz \xi^2\,.
\esp
\ee
It is useful to distinguish the cases of $\xi \to \infty$ 
(heavy $Z'$) and $\xi \to 0$ (light $Z'$). In the case of 
a heavy $Z'$ boson Eq.~\eqref{eq:kappa/tau} implies that
\be
\frac{\xi}{\tan\beta}\simeq \frac{\sz \xi^2}{z_N \mathscr{Z}}
\label{eq:largexi}
\ee
for small values of $|\sz|$.
In Eq.~\eqref{eq:largexi} we introduced the effective charge ratio
\be 
\label{eq:x-def}
\mathscr{Z}(\mu) = \frac{z_\phi(\mu)}{z_N}
\ee
that contains all dependence on the specific U(1) extension. 
As the effective charge of the BEH field depends on the 
renormalization scale, so does $\mathscr{Z}$, which we suppress 
in the following, but take into account as theoretical uncertainty 
of our predictions as discussed in App.~\ref{app:etarunning}.

Then in the limit of small neutral gauge mixing and heavy $Z'$, 
the V-A couplings simplify to 
\be 
\bsp
v_{Z',\nu} \simeq 
\sz\xi^2 \biggl(\frac{1}{\mathscr{Z}}-\frac{1}{2}\biggr), 
&\quad
a_{Z',\nu} \simeq  -\frac{1}{2}\sz \xi^2,
\\
v_{Z',\ell} \simeq 
\sz\xi^2 \biggl(\frac{1}{\mathscr{Z}}-\frac{3}{2}\biggr), 
&\quad
a_{Z',\ell} \simeq \frac{1}{2}\sz \xi^2,
\\
v_{Z',u} \simeq 
- \frac{\sz\xi^2}3 \biggl(\frac{1}{\mathscr{Z}}-\frac{5}{2}\biggr), 
&\quad
a_{Z',u} \simeq  -\frac{1}{2}\sz \xi^2,
\\
v_{Z',d} \simeq 
-\frac{\sz\xi^2}3\biggl(\frac{1}{\mathscr{Z}}+\frac{1}{2}\biggr),  
&\quad
a_{Z',d} \simeq   \frac{1}{2}\sz \xi^2
\,.
\label{eq:VAheavy}
\esp
\ee
As for a light $Z'$, Eq.~\eqref{eq:kappa/tau} implies
\be
\frac{\xi}{\tan\beta}\simeq -\frac{\sz}{z_N \mathscr{Z}}
\,,
\ee
and for the V-A couplings one has negligible $a_{Z',f}$, and 
\be 
\bsp
v_{Z',\nu} \simeq -\frac{\sz}{\mathscr{Z}}, 
&\quad
v_{Z',\ell} \simeq 
\sz\biggl(-\frac{1}{\mathscr{Z}}+2\cw^2\biggr), 
\\
v_{Z',u} \simeq 
\frac{\sz}{3} \biggl(\frac{1}{\mathscr{Z}}-4\cw^2\biggr), 
&\quad
v_{Z',d} \simeq 
\frac{\sz}{3}\biggl(\frac{1}{\mathscr{Z}}+2\cw^2\biggr)
\,.
\label{eq:VAlight}
\esp
\ee

We may also use the new gauge couplings as input parameters. To write the V-A couplings as functions of $g_z$, we first observe that
Eq.~\eqref{eq:mzpthetaz} implies
\be 
\bsp
\kappa & \simeq -\sz \quad\text{for}\quad \xi \to 0,
\\
\kappa & \simeq \sz \xi^2 \quad\text{for}\quad \xi \to \infty.
\esp
\ee
Then, for a heavy $Z'$ the axial couplings are 
$a_{Z',f} \simeq z_N \frac{2 g_z}{g_{Z^0}} a_f^{(h)}$
and $v_{Z',f} \simeq z_N \frac{2 g_z}{g_{Z^0}} v_f^{(h)}$.
Using the definition of $\kappa$ in Eq.~\eqref{eq:kappa-tau}, 
we obtain 
\be
a_\nu^{(h)} = a_u^{(h)} = -\frac14\,,\quad
a_\ell^{(h)} = a_d^{(h)} = +\frac14\,,
\ee
while from \eqref{eq:VAheavy}
\be 
\bsp
v_\nu^{(h)}&= 1 - \frac12\mathscr{Z}\,,
\quad\:
v_\ell^{(h)} = 1 - \frac32\mathscr{Z}\,,
\\
v_u^{(h)} &= -\frac13 + \frac56\mathscr{Z}\,,
\;
v_d^{(h)} = -\frac13\ - \frac16 \mathscr{Z}\,.
\esp
\ee
A light $Z'$ boson implies $a_{Z',f} \simeq 0$
and $v_{Z',f} \simeq z_N \frac{2 g_z}{g_{Z^0}} v_f^{(\ell)}$ where
\be
\bsp
v_\nu^{(\ell)}  &= 1\, \,,
\qquad\qquad\quad\;\;\;
v_\ell^{(\ell)}  = 1 - 2 \cw^2 \mathscr{Z}\,,
\\
v_u^{(\ell)}  &= -\frac13\ + \frac43 \cw^2 \mathscr{Z} \,,
\;
v_d^{(\ell)} = - \frac13 - \frac23 \cw^2 \mathscr{Z}\,.
\label{eq:vlight-cps}
\esp
\ee

\section{Direct Z-prime boson searches}
\label{sect:direct-zp-search}
Collider experiments such as LEP, Tevatron and the LHC performed direct searches for a $Z'$ boson. The nonobservation of such a particle can be and was translated into exclusion bands for the parameters of certain models predicting a $Z'$ boson.

According to Ref.~\cite{Appelquist:2002mw}, the results of the LEPII experiment show 
that $Z'$ is either heavier than the largest center-of-mass energy ($209~\GeV$) or 
$|z_\ell g_z| \lesssim 10^{-3}$. Tevatron searched for a $Z'$ in the range 
$200 ~\GeV < M_{Z'} < 800 \GeV$ \cite{Carena:2004xs}. Finally, ATLAS \cite{ATLAS:2019erb} and CMS \cite{CMS:2021ctt} at the LHC performed the most 
recent searches up to $M_{Z'} < 5500 \GeV$. Below $M_Z$, the NA64 and BaBar experiments together with FASER below the mass of the pion provide strong 
bounds for a light $Z'$ boson \cite{NA64:2019imj,BaBar:2017tiz}. In this study we focus on the
exclusion bounds obtained from ATLAS and CMS for a heavy, and NA64, BaBar and FASER for a light $Z'$.

In order to compare model predictions to experimental results at colliders
one has to compute the cross section for the process 
$p p \to Z' + X \to \ell^+ \ell^- + X$, which is usually performed in the narrow 
width approximation
\be 
\bsp
\label{eq:zp_prod1}
\sigma\bigl(pp &\to Z' + X \to \ell^+ \ell^- + X \bigr) = 
\\&=
\sigma\bigl(pp \to Z' X\bigr) \mathrm{Br}\bigl(Z' \to \ell^+ \ell^- \bigr),
\esp
\ee
assuming that the total width of the $Z'$ boson $\Gamma_{Z'}$ is much smaller than 
its mass, $\gamma_{Z'} = \Gamma_{Z'}/M_{Z'} \ll 1$. Eq.~\eqref{eq:zp_prod1} is 
usually presented as 
\be 
\bsp
\label{eq:zp_prod2}
\sigma\bigl(pp &\to Z' + X \to \ell^+ \ell^- + X\bigr) = 
\\&=
\frac{\pi}{6s}\biggl(c_{U} w_{U}\bigl(s, M_{Z'}\bigr)+c_{D} w_{D}\bigl(s, M_{Z'}\bigr)\biggr),
\esp
\ee
where and $U \in \{u,c,t\}$, $D \in \{d,s,b\}$ and the coefficients $c_q$ collect model dependent contributions to the cross section 
\be 
\label{eq:zp_cx_coupling}
c_q = \bigl(2 \sqrt{2} G_F M_Z^2 \rho\bigr)  
\mathrm{Br}\bigl(Z' \to \ell^+ \ell^- \bigr) 
\bigl ( a_{Z',q}^2 + v_{Z',q}^2\bigr)
\,.
\ee

The hadronic structure functions $w_{U/D}$ (cf.~Ref.~\cite{Carena:2004xs}) 
collect the QCD corrections. 
For the production of a heavy neutral gauge boson they depend only 
on the $M$ of the gauge boson and the center of 
mass energy squared $s$, 
\be
\bsp
w_{U/D} &= \sum_{q \in U/D}\int_0^1 \!\mathrm{d}x_1 
\int_0^1 \!\mathrm{d}x_2 \int_0^1 \!\mathrm{d}z 
~\delta\biggl(\frac{M^2}{s} - z ~ x_1 x_2 \biggr)
\\
&\times 
\biggl[
f_g\bigl(x_1,M\bigr)\biggl( 
f_q\bigl(x_2,M\bigr)+f_{\overline{q}}\bigl(x_2,M\bigr)
\biggr)\Delta_{gq}\bigl(z,M^2\bigr)
\\
&+
\biggl( 
f_q\bigl(x_1,M\bigr)f_{\overline{q}}\bigl(x_2,M\bigr)
\biggr)\Delta_{qq}\bigl(z,M^2\bigr) + \big( x_1 	\leftrightarrow x_2 \bigr)
\biggr],
\label{eq:wUD}
\esp
\ee
where the functions $f_i(x, \mu_F)$ are the parton distribution functions 
inside the proton for parton $i$ at factorization scale $\mu_F$. 
We use the NNPDF3.0 NLO PDF set in our 
numerical computations. At the next-to-leading order (NLO) accuracy 
the coefficient functions $\Delta_{ab}$ for vector boson production are 
known \cite{Altarelli:1979ub}. One also needs to compute the total decay width of the 
$Z'$ to obtain the cross section \eqref{eq:zp_prod1}. We collect the 
coefficient functions and the decay width formulae of the $Z'$ boson,
needed to compute the cross section in Eq.~\eqref{eq:zp_prod2} in App.~\ref{app:zp_decay}.

\section{Numerical analysis}
\subsection{Parameter scanning}
\label{sect:param-scan}
The model predictions can be expressed as functions of the free 
parameters of the theory. At the most fundamental level, these 
are the  free $z$ charges, new couplings and VEV ratio,
\be 
z_\phi\,,\: z_N\,,\: g_z\,,\: g_{yz}\,,\: \tan\beta\,,
\ee
which are not independent and a certain combination of them appears in the model predictions. For instance, the tree level $\rho$ parameter estimates the constraints  from the electroweak precision observables and it depends only on 
\be 
\label{eq:paramset-def}
\bigl(\sz\,,\: M_{Z'}\bigr) \quad\text{or}\quad
\bigl(z_N\,g_z \,,\: M_{Z'}\,,\: \mathscr{Z}\bigr)
\,.
\ee
The NA64 experiment presents exclusion bounds for a dark photon 
in the $(\epsilon,M_{A'})$ plane. Those constraints can be translated to our model parameters 
using the relation derived in App.~\ref{app:na64_match}. 
This shows one that exclusion bounds depend on either
\be
\label{eq:scanparam-set}
\bigl(\sz\,,\: M_{Z'}\,,\: \mathscr{Z}\bigr) \quad\text{or}\quad \bigl(z_N\,g_z \,,\: M_{Z'}\,,\: \mathscr{Z}\bigr),
\ee
Presently, the most stringent bounds on the parameter space for heavy $Z'$ 
bosons can be obtained from direct searches using the Drell-Yan pair production process 
\be 
p + p \to Z' + X \to \ell^+ +\ell^- + X,
\ee
described in Sect.~\ref{sect:direct-zp-search}. The corresponding cross section \eqref{eq:zp_prod2}  can be rewritten as 
\be
\bsp
\label{eq:scan_cx}
\sigma &= \frac{4\pi^2}{3 s}\frac{\Gamma_{Z'}}{M_{Z'}}\mathrm{Br}\bigl(Z' \to \ell^+ \ell^- \bigr)
\\&\times
\biggl(\mathrm{Br}\bigl(Z' \to U \overline{U} \bigr) w_{U}\bigl(s, M_{Z'}\bigr)
\\&\quad+
\mathrm{Br}\bigl(Z' \to D \overline{D} \bigr)
w_{D}\bigl(s, M_{Z'}\bigr)\biggr),
\esp
\ee
where the branching fractions are listed in App.~\ref{app:zp_decay},
while $w_{U/D}$ are given by Eq.~\eqref{eq:wUD}. In this case as well, the predictions depend on the parameter set \eqref{eq:scanparam-set} or equivalently on
\be 
\bigl(\gamma_{Z'}, M_{Z'}, \mathscr{Z}\bigr)
\ee
where $\gamma_{Z'} = \Gamma_{Z'}/M_{Z'}$.

\subsection{Constraints on a light neutral gauge boson}
Light vector-type particles, usually called dark photons $(A')$, are often considered as a portal to a secluded sector in particle physics or downright as dark matter candidates. 
Presently, the most stringent, $90\%$ CL exclusion bound in the dark photon mass range $M_{A'} \in (1~\MeV,8~\GeV)$ comes from the combined results of the NA64 \cite{NA64:2019imj}, BaBar \cite{BaBar:2017tiz} and more recently the FASER \cite{FASER:2023tle} experiments. 
The dark photon model probed in these experiments have a single vector type coupling  $\epsilon e$ to the electromagnetic current. 
The parameters $\bigl(\epsilon,M_{A'}\bigr)$ can be matched to a generic U(1)$_z$ as detailed in App.~\ref{app:na64_match}.
We scanned the parameter planes $\bigl(\sz, M_{Z'}\bigr)$ and 
$\bigl( |z_N\,g_z |, M_{Z'} \bigr)$ for several benchmark values 
of $\mathscr{Z}$. Using the value for the $\rho$ parameter given 
in Eq.~\eqref{eq:u1_rho_exp} we set upper bounds 
\be 
\label{eq:rho-light-bound}
|\sz| \lesssim 4.5 \cdot 10^{-3} 
\quad\text{and}\quad
|z_N\,g_z | \lesssim \frac{1.7 \cdot 10^{-3} }{|\mathscr{Z}|}
\ee
for $M_{Z'} \ll M_Z$.

The experimental bounds obtained from NA64, BaBar and FASER all 
depend on $v_\ell^{\ell}$ given in Eq.~\eqref{eq:vlight-cps}. 
The former two experiments searched for invisible decay products 
of dark photons, whereas the latter one searches for decays 
$A' \to e^+ e^-$ which introduces further dependence on the
corresponding branching fractions. The mapping of $\epsilon$ 
onto $\sz$ or $|z_N\,g_z|$ and $M_{Z'}$ thus leads to a 
dependence on $\mathscr{Z}$ in the exclusion bands. For instance,
as $\mathscr{Z}$ approaches $1/(2\cw^2)$, the reduced vector 
coupling tends to zero $v_\ell^{\ell} \to 0$, which renders 
$|\sz|$ and $|z_N\,g_z |$ unconstrained. The exclusion band 
obtained from the FASER experiment is even more sensitive as 
the branching fraction $\mathrm{Br}(Z' \to e^+ e^-)$ also depends
on $v_\ell^{\ell}$.

\begin{figure}[t]
\includegraphics[width=0.4\linewidth]{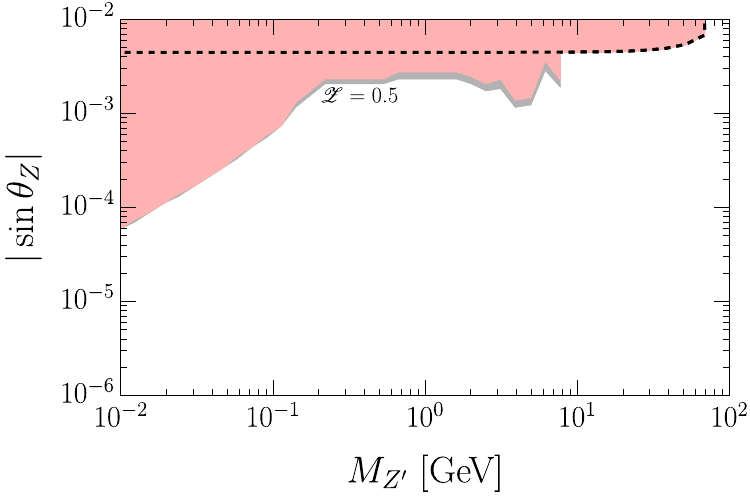}
\includegraphics[width=0.4\linewidth]{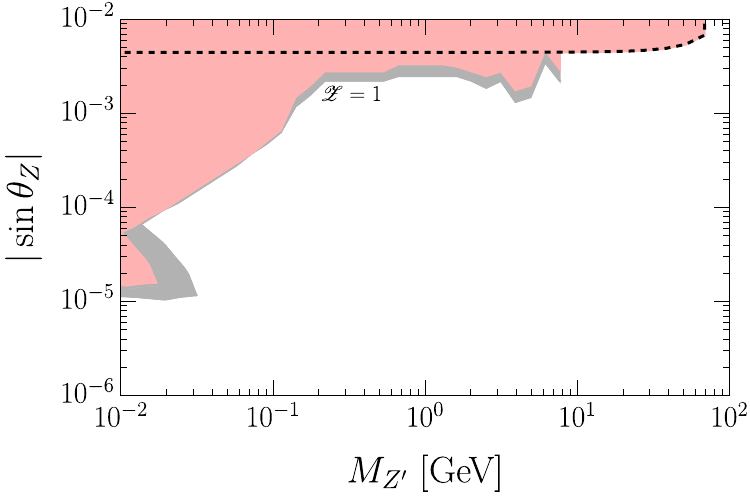}
\\
\includegraphics[width=0.4\linewidth]{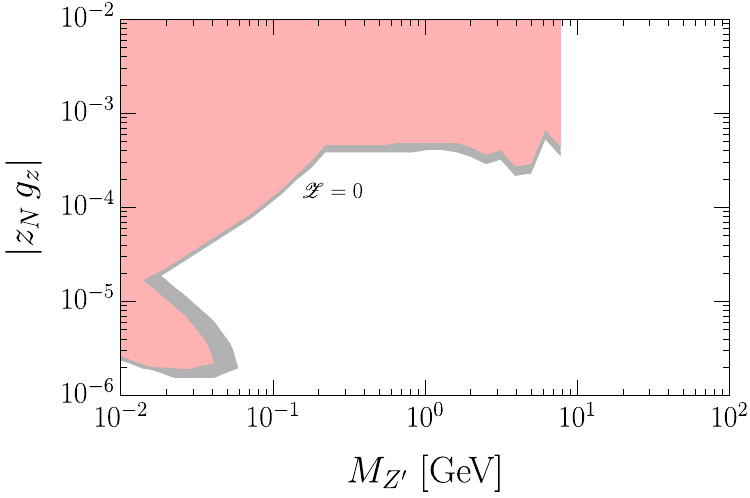}
\includegraphics[width=0.4\linewidth]{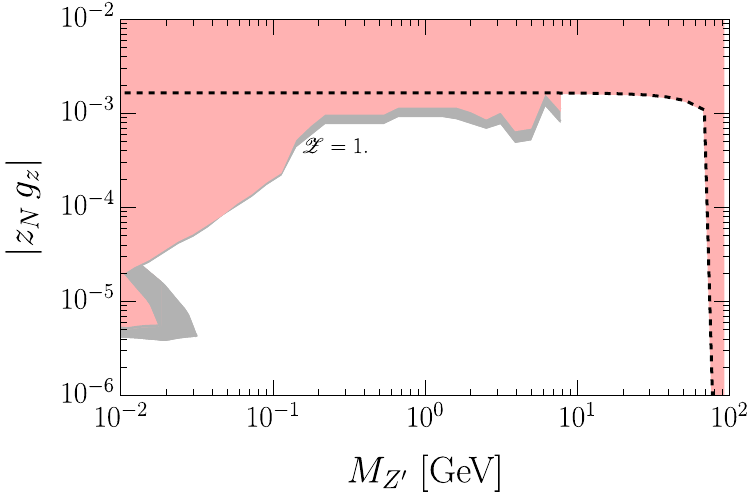}
\caption{\label{fig:na64-bands} 
$90 \%$ CL exclusion bounds for light $Z'$ bosons obtained from the NA64, BaBar and FASER experiments. The width of the band corresponds to the uncertainty in the number of
sterile neutrino families where the decay $Z \to N + N$ is kinematically allowed. The region above the dashed line is excluded due to the $\rho$ parameter and the area above the gray bands is also excluded for a selected value of $\mathscr{Z}$.
}
\end{figure}
Our findings for selected benchmark values of $\mathscr{Z}$ are summarized in Fig.~\ref{fig:na64-bands}. The regions in 
the parameter planes above the dashed line and gray bands are excluded 
at $90 \%$ CL. The dashed lines correspond to the experimental value of the $\rho$ parameter in Eq.~\eqref{eq:u1_rho_exp},
whereas the regions above the gray bands correspond to the  exclusions 
by direct searches at fixed values of the effective charge ratio $\mathscr{Z}$. 
The width of the gray bands is the uncertainty due to the number of 
right handed neutrinos lighter than $M_{Z'}/2$. A light $Z'$ boson 
may always decay into the three families of active neutrinos, but 
decays into right handed neutrinos may be kinematically forbidden depending on 
the specific values of $M_{Z'}$. 

\subsection{Constraints on a heavy neutral gauge boson}

\begin{figure}[t]
{
\includegraphics[width=0.4\linewidth]{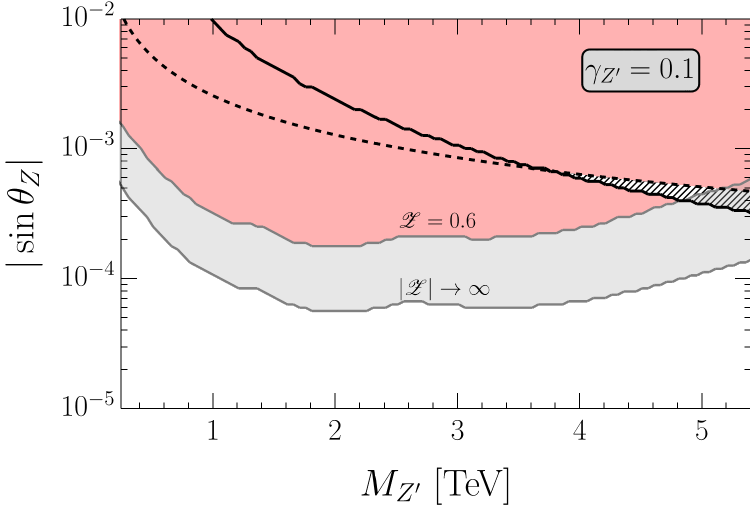}
\includegraphics[width=0.4\linewidth]{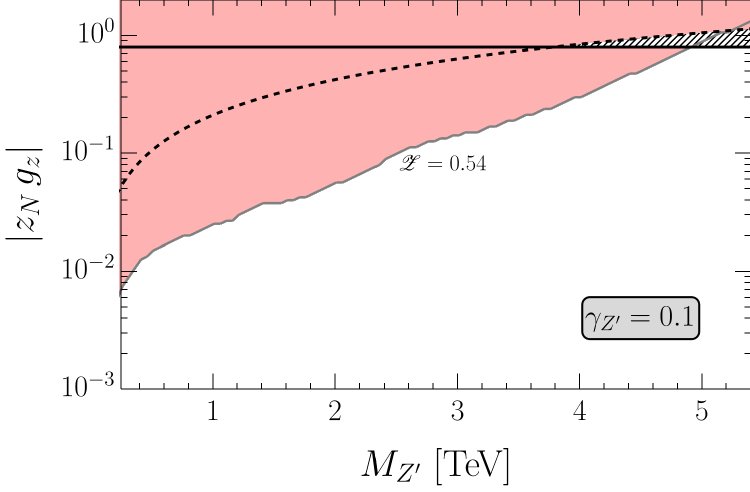}
}
\\
{
\includegraphics[width=0.4\linewidth]{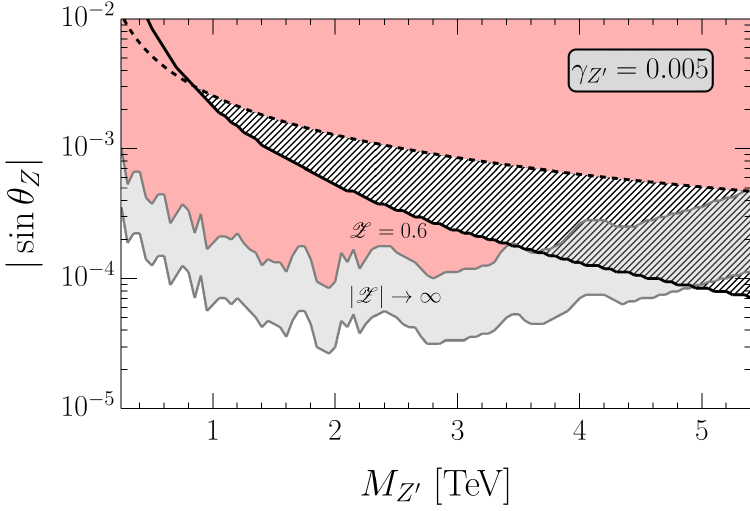}
\includegraphics[width=0.4\linewidth]{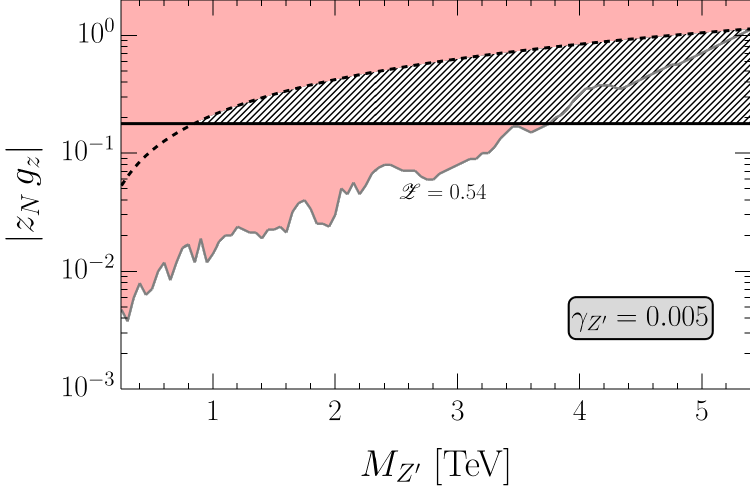}
}
\caption{\label{fig:lhc-bands} 
$95 \%$ CL exclusion bounds for heavy $Z'$ bosons obtained from the CMS and ATLAS experiments at the LHC for fixed ratios $\gamma_{Z'}$. The region above the dashed line is excluded due to the $\rho$ parameter and the area above the gray line is also excluded for a select value of $\mathscr{Z}$.
}
\end{figure}

Direct searches for heavy $Z'$ bosons were performed at the LEPII, Tevatron and LHC as well and are of continued interest for future colliders \cite{FCC:2018bvk,Mangano:2017tke}.
We perform the scan in the parameter sets given in Eq.~\eqref{eq:scanparam-set} using the 95 $\%$ CL exclusion bands presented by the ATLAS \cite{ATLAS:2019erb} and CMS \cite{CMS:2021ctt} experiments. Our findings are summarized in Figs.~\ref{fig:lhc-bands}. The exclusion limit by the $\rho$ parameter is represented again with a dashed line: the region above it is excluded. Analytically this correspond to
\be 
|\sz| \lesssim 0.0025\biggl[\frac{1~\TeV}{M_{Z'}}\biggr]
\quad\text{and}\quad
|g_z \, z_N| \lesssim \frac{0.11}{\mathscr{Z}}\biggl[\frac{M_{Z'}}{1~\TeV}\biggr].
\ee

The collider searches by ATLAS and CMS are performed for fixed values 
of the ratio $\gamma_{Z'}$. We chose the datasets corresponding to 
the largest $(\gamma_{Z'} = 10\%)$ and smallest  
$(\gamma_{Z'} = 0.5\%)$ presented values. 
It is possible that the cross section \eqref{eq:scan_cx} is large 
enough so that the process is excluded experimentally for 
given values of the input parameters \eqref{eq:scanparam-set}, 
but the corresponding ratio $\gamma_{Z'}$ is larger than that 
searched for in the experiment in a region whose lower boundary 
is denoted by a solid curve in Figs.~\ref{fig:lhc-bands}. 
The hatched region corresponds to this exact case, i.e.~where no 
strict exclusion applies. The region shaded in red in the parameter 
plane presented in Figs.~\ref{fig:lhc-bands} is excluded at 95\,\% CL. 
The branching fractions and the ratio $\gamma_{Z'}$ in 
Eq.~\eqref{eq:scan_cx} depend on the charge ratio $\mathscr{Z}$,
hence do so the exclusion bounds. We find that there exist a 
value of $\mathscr{Z}$ both for $\sz$ and $z_N\,g_z $ which 
corresponds to a loosest, i.e.~the most conservative bound on 
these parameters. For the mixing $\sz$ this value is $\mathscr{Z} \simeq 0.6$.
Any other fixed $\mathscr{Z}$ value presents a more severe bound than that shown
in Fig.~\ref{fig:lhc-bands}. It is interesting that the cross section in Eq.~\eqref{eq:scan_cx} 
diverges as $|\mathscr{Z}| \to 0$ (vanishing $z_N$ charge), which means that only zero mixing $\sz = 0$ is 
allowed for small $\mathscr{Z}$. Conversely, Eq.~\eqref{eq:scan_cx} saturates at 
a finite value for $|\mathscr{Z}| \to \infty$, the corresponding exclusion bands 
are shown in the left hand side plots of Fig.~\ref{fig:lhc-bands}.

As for $z_N\,g_z $ the most conservative bound corresponds to  
$\mathscr{Z} \simeq 0.54$ as can be seen in the plots on the right hand 
side of Fig.~\ref{fig:lhc-bands}. As opposed to the exclusion
bound on $\sz$, in this case the cross section is finite at $\mathscr{Z}=0$,
hence one has a well defined exclusion bound on $z_N\,g_z $ for $\mathscr{Z} =0$.

\subsection{Constraints on the parameter space of specific U(1) extensions}

We showcase the exclusion bounds on two specific U(1)$_z$ extensions, 
one with a light $Z'$ boson and one with a heavy $Z'$ boson. We consider here the uncertainty due to the RGE running of $\eta$ and we also use the mass of the $W$ boson as a constraint.

The tree level $\rho$ parameter discussed in 
Sect.~\ref{subsect:mod-rho} is a useful quantity to gauge the 
exclusion from electroweak precision observables in a model independent manner. 
The effect of one-loop BSM corrections might become important for a given region in the parameter space and thus the use of a precise prediction is warranted. 
The drawback is that the radiative BSM corrections are in 
general complicated functions of the free parameters and the $z$ charges. 
Once the $z$ charges are set and $\eta$ is considered as an uncertainty, there are two free parameters from the gauge sector ($M_{Z'}$ and either $\sz$ or $g_z$) 
and two from the scalar sector ($M_S$ and $\sss$). 
Using these four parameters we compute the complete one-loop 
corrections to $M_W$ in U(1)$_z$ extensions presented in Ref.~\cite{Peli:2023fyb} 
based on the computational method of Ref.~\cite{Athron:2022isz},
consult also Ref.~\cite{Peli:2023fxw} for the renormalization of $\sz$.
Our input parameters are 
\be 
M_W^{\text{SM}} = 80.353~\GeV,
\quad
M_W^{\text{exp.}} = 80.377~\GeV,
\ee
with a combined experimental and theoretical uncertainty of $\sigma = 15\,\MeV$. 

The BSM corrections can either amount to a positive or negative contribution to $M_W$. 
A heavy $Z'$ boson and a light $S$ scalar ($M_S < M_H$) increase, 
while a light $Z'$ boson and a heavy $S$ scalar ($M_S > M_H$) decrease the predicted value of $M_W$.
In this work we focus on the effect of the gauge sector and thus we present exclusion bounds obtained from $M_W$ at $\sss = 0$, i.e. when the extended scalar sector does not affect the mass of the $W$ boson.

Our case study for a U(1)$_z$ extension with a light $Z'$ boson is 
the SWSM (recall that $z_N=1/2$ and $z_\phi = 1$). This model can 
explain the observed dark matter abundance in the Universe with 
freeze-out scenario if 
$10\,\MeV \lesssim M_{Z'}  \lesssim  m_{\pi}\ll M_Z$ and the dark 
matter candidate is the lightest sterile neutrino, which is 
considered to be lighter than $M_{Z'}/2$, while the other sterile
neutrinos are much heavier~\cite{Iwamoto:2021fup}.

\begin{figure}[t]
{
\includegraphics[width=0.4\linewidth]{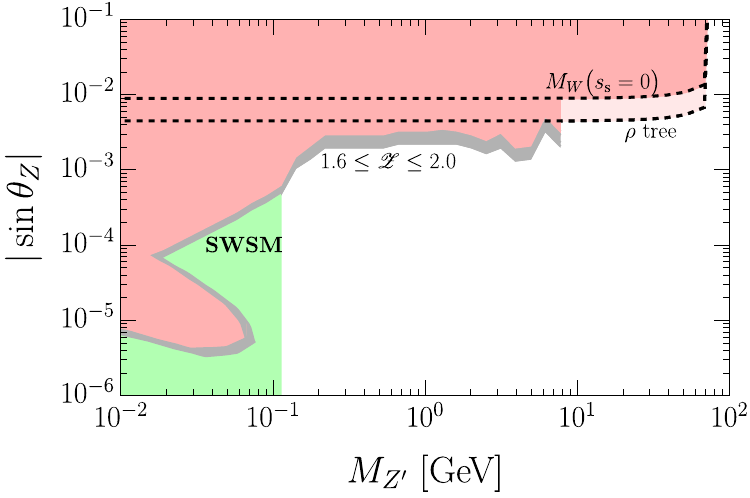}
 \includegraphics[width=0.4\linewidth]{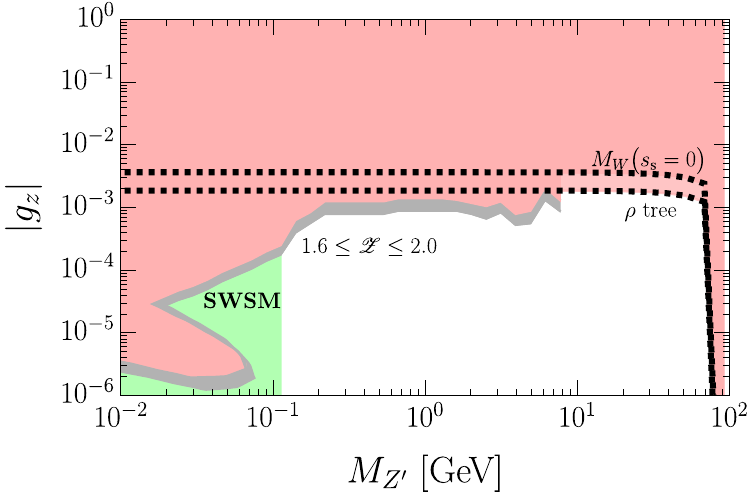}
}
\caption{\label{fig:swsm_bound} 
Exclusion bounds for models with a light $M_{Z'}$ and $z_N=1/2$ and $z_\phi = 1$. 
The red region is excluded at $90~\%$ CL. 
The green region is the preferred parameter space of the SWSM.
The width of the lines take into account the uncertainty in the $\eta$ parameter. 
The gray line corresponds to the NA64, BaBar and FASER experiments whereas the dashed ones correspond
to the bounds from $M_W$ and $\rho$.
}
\end{figure}
Our findings are summarized in Fig.~\ref{fig:swsm_bound}. The region 
in red is excluded at $90~\%$ CL. The gray band is the lower boundary 
of the exclusion region from the NA64, BaBar and FASER experiments and the 
width of the gray band corresponds to the combined uncertainty from
decays of the $Z'$ boson and the running of $\eta$. Solving the RGEs 
of App.~\ref{app:etarunning}, we find that the largest possible 
value of $\eta$ is $0.4$, hence $\mathscr{Z}$ can take values in the 
range $(1.6,2.0)$. The dashed lines correspond to the bound from 
$M_W$ computed with $\sss = 0$ and to the bound from the tree level 
$\rho$ parameter as a reference. The scalar sector has the potential 
to significantly affect the bound obtained from $M_W$. In fact, for a heavy scalar ($M_S \gg M_h$) and a light $Z'$ boson one may
write the BSM correction $\delta M_W^{\text{BSM}}$ to the mass of the $W$ boson as 
\be 
\delta M_W^{\text{BSM}}\simeq - \left[5.6 \big(100 \sz\,\big)^2
+ 1.5\big(10\,\sss\big)^2 \biggl(1 + 0.57 \log\biggl( \frac{M_S}{1~\TeV}\biggr) \biggr)\right]\MeV,
\ee
which is independent of the $z$ charges. For instance, a scalar with a mass $M_S \simeq 1~\TeV$ and a mixing of $\sss \simeq 0.2$ would increase the difference $|M_W - M_W^{\text{exp.}}|$ above $2\sigma$, excluding any nonzero value of $\sz$.

\begin{figure}[t]
\includegraphics[width=0.4\linewidth]{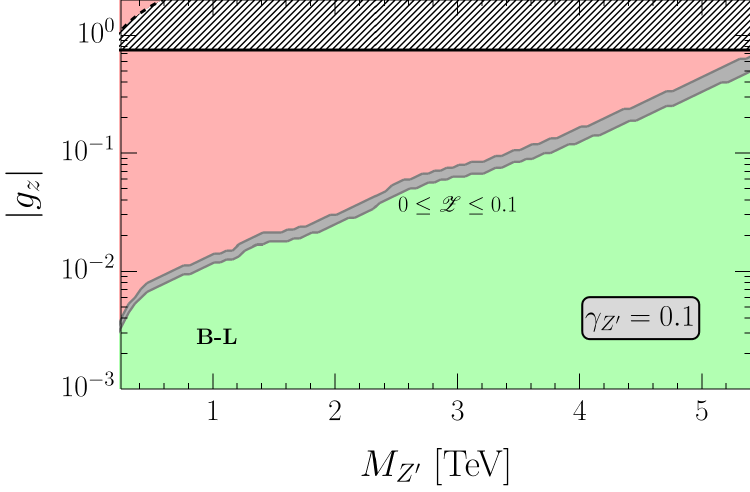}
 \includegraphics[width=0.4\linewidth]{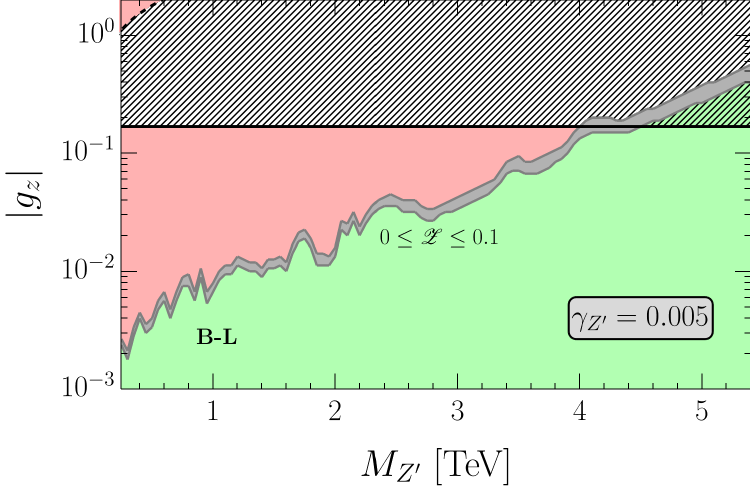}
\caption{\label{fig:b-l_bound} 
Exclusion bounds for the $B-L$ extension of the SM where one has $z_N=1/2$ and $z_\phi = 0$. 
The red region is excluded at $95~\%$ CL. 
The green region is the preferred parameter space of the $B-L$ model.
The width of the lines take into account the uncertainty in the $\eta$ parameter. 
The gray line corresponds to the CMS and ATLAS direct searches at $\gamma_{Z'} = 0.1$ and at $0.005$. 
The dashed line in the top left corner corresponds to the exclusion from $M_W$ at $\sss = 0$.
}
\end{figure}
As for a U(1)$_z$ extension with a heavy $Z'$ boson, $M_{Z}\ll M_{Z'}$, we choose to investigate the  $B-L$ extension of the SM, 
which has $z_N=1/2$ and $z_\phi = 0$. 
It is interesting to note that in this case $\mathscr{Z} =0$ at the default scale $\mu_0$, and the uncertainty from the RG running of $\eta$ is also essentially negligible, at most about 0.1. 
Hence, there is effectively no mixing between the $Z$ and $Z'$ bosons, 
$\sz \simeq 0$, and consequently, there is no bound from the tree level $\rho$ parameter.
Our findings are summarized in Fig.~\ref{fig:b-l_bound}. The region in red is excluded at $95~\%$ CL. 
The dashed line correspond to the exclusion from $M_W$ at $\sss = 0$. Since the tree level $\rho$ parameter equals one, this corresponds purely to one-loop BSM corrections to $M_W$.
The hatched region is not excluded by $M_W$ and the width to mass ratio $\gamma_{Z'}$ of the $Z'$ boson is larger than the one considered experimentally. The green region displays the presently allowed parameter space of the $B-L$ model. 

\subsection{Projections for future pp collider experiments}

\begin{figure}[t]
{
\includegraphics[width=0.4\linewidth]{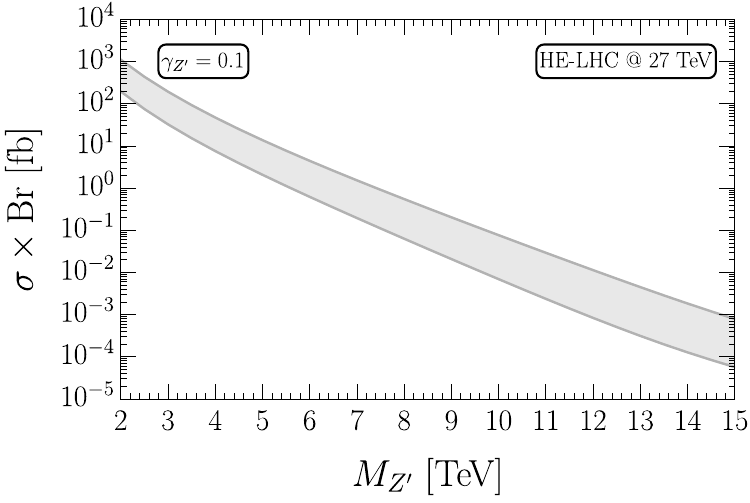}
\includegraphics[width=0.4\linewidth]{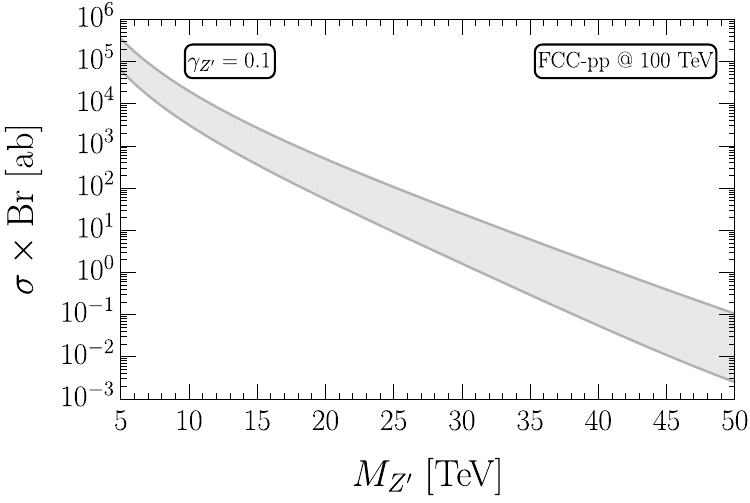}
}
\caption{\label{fig:pp_future_cx} 
Production cross sections $\sigma\bigl( pp \to Z'\bigr)$ times the leptonic branching fraction $\text{Br}\bigl(Z' \to \ell^+ + \ell^-\bigr)$ as the function of $M_Z'$ for center of mass energies $\sqrt{s} = 27~\TeV$ (left, for the HE-LHC) and $100~\TeV$ (right, for FCC-pp) and fixed ratio $\gamma_{Z'} = 0.1$.
The cross section for any value of $\mathscr{Z}$ is inside of the gray band.
}
\end{figure}

The High Energy LHC experiment is planned to operate at $\sqrt{s} = 
27\,\TeV$ center of mass energy, while the Future Circular Collider 
will collide particles at $\sqrt{s} = 100\,\TeV$. The experimental 
programs of both machines include direct searches for $Z'$ 
bosons. The cross section \eqref{eq:scan_cx} for the process 
$p + p \to Z' + X \to \ell^+ +\ell^- + X$, which is the main search
channel for $Z'$ bosons, is shown on Fig.~\ref{fig:pp_future_cx} 
for relevant values of $\sqrt{s}$. The cross sections are inside 
of the gray band on Fig.~\ref{fig:pp_future_cx} for any value of 
$\mathscr{Z}$ for $\gamma_{Z'} = 0.1$. Note that the gray band for a different 
value of $\gamma_{Z'}$ can be obtained by the linear rescaling of 
those on Fig.~\ref{fig:pp_future_cx}.

\begin{figure}[t]
{
\includegraphics[width=0.4\linewidth]{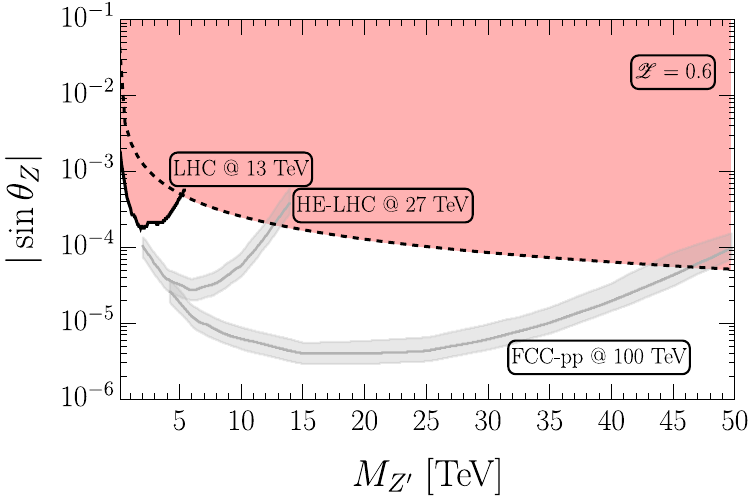}
 \includegraphics[width=0.4\linewidth]{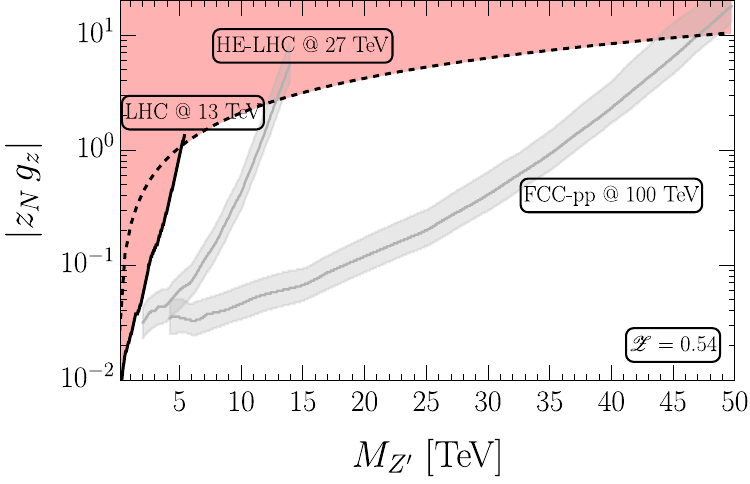}
}
\caption{\label{fig:pp_future} 
Projected exclusion bounds on $|\sz|$ at $\mathscr{Z}=0.6$ and 
 on $| z_N \, g_z |$ at $\mathscr{Z}=0.54$ for the HE-LHC and FCC-pp experiments using the simulated exclusion bands obtained in Refs.~\cite{FCC:2018bvk} and \cite{CERN-ACC-2019-028}.
The dark gray line is the expected median exclusion limit and the width of the gray bands correspond to the $95\%$ CL expected limit. The dashed line represents the exclusion by the $\rho$ parameter and the red shaded area is already excluded at $95\%$ CL.
}
\end{figure}

Detector simulations are already available both for the HE-LHC 
and the FCC-hh. We compute projected 95 $\%$ CL exclusion bands 
both for $|\sz|$ and $|g_z \, z_N|$ using the simulations
for the HE-LHC at $15$ ab$^{-1}$ integrated luminosity 
\cite{FCC:2018bvk} and for the FCC-hh  at $30$ ab$^{-1}$ 
\cite{CERN-ACC-2019-028}. Our predictions are shown in Fig.~\ref{fig:pp_future}. 
The width of the gray bands correspond to the 2\,$\sigma$ 
uncertainty of the simulation in the location of the exclusion band.

It is noteworthy that for large $\bigl( \gtrsim 10~\TeV \bigr)$ masses of the $Z'$ boson the cross section
\be
\sigma\bigl(p + p \to Z' + X \to Z + W^+ + W^- + X\bigr)
\ee
may become comparable to or larger than the Drell-Yan 
pair production cross section in Eq.~\eqref{eq:scan_cx} as the ratio 
of the two cross sections is
\be
\bsp
\frac{\mathrm{Br}\bigl(Z' \to Z W^+ W^-\bigr)}{\mathrm{Br}\bigl(Z' \to \ell^+\ell^-\bigr)} &= \frac{\mathscr{Z}^2}{2-6 \mathscr{Z} + 5 \mathscr{Z}^2} \biggl( C_{ff}\frac{7 \cw^4} {160\pi}\biggr)\frac{M_{Z'}^2}{M_Z^2}
\\
&\simeq 0.368 ~ \frac{\mathscr{Z}^2}{2-6 \mathscr{Z} + 5 \mathscr{Z}^2} \biggl[\frac{M_{Z'}}{10 ~\TeV}\biggr]^2.
\esp
\ee

\section{Conclusions}

In this paper we studied the parameter space of U(1)$_z$ 
extensions of the standard model with an additional complex scalar field 
and three families of right handed neutrinos with generation independent
$z$ charges and no exotic fermions. Anomaly cancellation 
constrains the $z$ charges such that two $z$ charges remain arbitrary. 
The vector - axial vector couplings, which are critical in the analysis 
presented in this work, depend on a special combination $\mathscr{Z}$ of $z_\phi$ 
the $z$ charge of the BEH field and $z_N$ as given in Eq.~\eqref{eq:x-def}.
We presented our predictions using $\mathscr{Z}$, the mass $M_{Z'}$ of the $Z'$ boson 
and either the $\sz$ mixing between $Z$ and $Z'$, or $z_N\,g_z $ the
right handed neutrino $z$ charge times the new gauge coupling.
Our exclusion bounds on these parameters used the results of the
NA64, BaBar, FASER experiments for a light and those of the  ATLAS and CMS 
experiments for a heavy $Z'$ boson as constraints from direct searches.
We also studied the limits obtained from the measured values of the 
$\rho$ parameter and the mass of the $W$ boson as constraints from indirect sources.

For a light $Z'$ boson the $\rho$ parameter provides a bound on $|\sz|$, 
independent of $\mathscr{Z}$, and a bound on $|z_N\,g_z |$, proportional to $1/\mathscr{Z}$ 
as given in Eq.~\eqref{eq:rho-light-bound}. The bound on $|\sz|$ obtained 
from the NA64, BaBar and FASER experiments is in general more severe,
and it is proportional to $1/\mathscr{Z}$. The constraint on $|z_N\,g_z |$ 
from such direct searches depends weekly on $\mathscr{Z}$ unless $\mathscr{Z}$ is fine tuned to $\mathscr{Z}=1/\bigl( 2 \cw^2\bigr)$, (see Eq.~\eqref{eq:match-eps-gz}).  

In the case of a heavy $Z'$ boson, the value $\mathscr{Z} \simeq 0.6$ 
corresponds to the loosest bound on $|\sz|$ (or $\simeq 0.54$ on $|z_N\,g_z|$)
obtained from the ATLAS and CMS experiments, which means that 
one has a model independent way to constrain the parameter space of U(1)$_z$
extensions with different charge assignments. In this region, the $\rho$ 
parameter excludes a decent portion of the parameter space, but it is never 
as severe as the exclusion bound obtained from direct searches.

We also used detector simulations for the HE-LHC and FCC experiments to 
provide projected exclusion bounds on the parameter space.  We found that 
the minimum of the excluded $|\sz|$ values will improve by one order of 
magnitude in HE-LHC compared to LHC and an additional one order of magnitude in FCC-hh 
compared to HE-LHC. 
As for $|g_z \, z_N|$ we find no such improvement.
The exclusion bound obtained from the LHC and the projected ones are based on leptonic final states.
We find that for very large values of $M_{Z'}$ $(>10~\TeV)$ the 
decay $Z' \to Z + W^+ + W^-$ might become dominant over 
$Z' \to \ell^+ + \ell^-$ depending on $\mathscr{Z}$. 
This means that we propose direct searches at FCC for the final 
state $Z + W^+ + W^-$ as it is well motivated experimentally. 

\subsection*{Acknowledgment}

We appreciate collaboration with Jos\'u Hern\'andez-Garc\'ia at 
early stages of this project. We are grateful to members of the 
PPPheno Group at ELTE for useful discussions and especially to 
K.~Seller for his  useful contributions. This research was supported by the 
Excellence Programme of the Hungarian Ministry of Culture and 
Innovation under contract TKP2021-NKTA-64
and by the Hungarian Scientific Research Fund PD-146527.

\appendix

\section{Running eta parameter}
\label{app:etarunning}

We defined the ratio of the mixing coupling $g_{yz}$ and the new gauge coupling $g_z$ as
\be 
\eta = \frac{g_{yz}}{g_z}
\,.
\ee
It always appears as a correction to the $z$ charge of the BEH field as in Eq.~\eqref{eq:zphi(mu)}. Like any other coupling, $\eta$ 
depends on the renormalization scale $\mu$, as described by the 
renormalization group equations (RGEs)
\be 
\bsp
\dot{g}_y &= \frac{g_y}{16 \pi^2} \biggl(\frac{41}{6}g_y^2 + \frac{5}{3}g_z^2 \, \eta \,b_\eta\biggr), 
\\
\dot{g}_z  &= \frac{g_z^3}{16 \pi^2}\biggl(\frac{5}{9} + \frac{4000}{369}z_N^2 + \frac{10}{41}b_\eta^2 \biggr),
\\
\dot{\eta} &= \frac{g_y^2}{16\pi^2} b_\eta
\label{eq:etaRGEs}
\esp
\ee
where
\be
b_\eta = \frac{16}{3}z_N -\frac{41}{3}\biggl(z_\phi  - \frac{\eta}{2}\biggr)
= \frac{z_N}{3}\bigl(16 - 41 \mathscr{Z}\bigr)
\ee
is a linear function of $\eta$.
The derivative $\dot{f} = \partial f / \partial t$ is meant with 
respect to $t = \ln\bigl(\mu / \mu_0\bigr)$. We chose
the $z$ charge assignment according to Tab.~\ref{tab:zchrges} 
at an arbitrarily chosen fixed 
scale $\mu_0$ where $\eta(\mu_0) =0$. Then one can investigate the 
uncertainty due to the choice of the unknown scale $\mu_0$. 

The scale $\mu_0$ can be chosen arbitrarily. In our study the most
reasonable choices are either $\mu_0 = m_{\rm t}$ or $\mu_0= M_{Z'}$. 
The values of $M_{Z'}$ considered here are at maximum a few 
tens of $\TeV$, then the running of $\eta$ from $\mu_0 = M_{Z'}$ 
down to the electroweak scale does not affect the value of $\mathscr{Z}$ in 
any way relevant for the phenomenology considered in this work. 
One may set $\mu_0$ as high as $\Mpl$, in which case $\eta({m_{\rm t}}) = \rm{O}(1)$, with exact value depending on $z_N$ and $z_\phi$.
For instance, in the SWSM, choosing   $\eta(\Mpl) =0$ at the 
Planck-scale, the renormalization group running with GUT normalization 
implies that  $\eta \simeq 0.67$ at the electroweak scale \cite{Iwamoto:2021fup}. 
Here we choose $\mu_0 = m_{\rm{t}}$. Then one initial condition, $g_y(m_{\rm{t}}) \simeq 0.36$ is known, 
while the initial conditions for $g_z$, $\eta$ as well as the $z$ 
charges $z_N$ and $z_\phi$ are free parameters.
One can show that the coupling $g_z(\mu)$ has a Landau pole below the 
Planck scale $\Mpl$ if $g_z(m_{\rm t})$ is larger than a critical value. Assuming a constant $\eta$, this value is
\be 
\alpha_z(m_{\rm t}) = \frac{g_z(m_{\rm t})^2}{4\pi} \gtrsim \frac{11.95}{41+800 z_N^2 + 18 b_\eta^2} 
\,.
\ee
Taking the running of $\eta$ into account, this formula is not exact and the 
actual upper bound on $g_z(m_{\rm t})$ to avoid the Landau pole below 
the Planck scale is about $15\%$ lower. The loosest constraint 
obtained from avoiding the Landau pole corresponds to 
$z_N = z_\phi = 0$ with $\eta(m_{\rm t} ) = 0$. Then one has the 
upper bound
$g_z(m_{\rm t}) \lesssim 1.91$. Any different $z$ charge assignment 
results in a considerably more severe upper limit. For instance,
in the SWSM it is $g_z(m_{\rm t}) \lesssim 0.22$.
Then the initial conditions $\eta(\Mpl) = 0$ and 
$\gz(m_{\rm t}) \in [0,0.22]$ in the one-loop RGEs of 
Eq.~\eqref{eq:etaRGEs} yield $\eta$ at the electroweak scale in the 
range $\eta(m_{\rm t}) \in [0.4,0.375]$ (the larger 
$\gz(m_{\rm t})$, the smaller $\eta(m_{\rm t})$).

\section{Chrial couplings of the Z and $Z'$ bosons}
\label{app:Chiralcouplings}

We list here the chiral couplings of the $Z'$ bosons to fermions in terms of 
the neutral mixing angle and effective couplings $\kappa$ and $\tau$ in Table 
\ref{tab:sw-chiral-cps-Zp}. The chiral couplings to the $Z$ boson can be obtained
by the replacement $(\cz,\:\sz) \to (\sz,\: -\cz)$.
\begin{table}[ht]
\centering
\begin{tabular}{ |c||c|c| } 
 \hline
 ~~~$f$~~~ & $C^R_{Z',f\overline{f}}$ & $C^L_{Z',f\overline{f}}$ \\ 
\hline\hline
$\nu$ &  
$\frac{\tau}{\tanb} z_N \cz$ 
& 
$-\sz + \bigl(-\kappa + \frac{\tau}{\tanb}z_N\bigr)\cz$ 
\\ [1ex] \hline
$\ell$ & 
$-2\sw^2 \sz + \bigl(-2\kappa + \frac{\tau}{\tanb}z_N \bigr)\cz$ 
& 
$\bigl(1-2\sw^2\bigr)\sz + \bigl( -\kappa + \frac{\tau}{\tanb}z_N\bigr)\cz$
\\ [1ex] \hline
$u$ & 
$\frac{4}{3}\sw^2\sz + \bigl(\frac{4}{3}\kappa - \frac{1}{3}\frac{\tau}{\tanb}z_N\bigr)\cz$ 
& 
$\bigl(-1+\frac{4}{3}\sw^2\bigr)\sz
+\frac{1}{3}\bigl(\kappa - \frac{\tau}{\tanb}z_N\bigr)\cz$ 
\\ [1ex] \hline
$d$ & 
$-\frac{2}{3}\sw^2\sz - \bigl(\frac{2}{3}\kappa + \frac{1}{3}\frac{\tau}{\tanb}z_N\bigr)\cz$ 
& 
$\bigl(1-\frac{2}{3}\sw^2\bigr)\sz
+\frac{1}{3}\bigl(\kappa - \frac{\tau}{\tanb}z_N\bigr)\cz$ 
\\ [1ex] \hline
\end{tabular}
\caption{The chiral couplings of the $Z'$ boson to fermions in U(1)$_z$ extensions of the SM.}
\label{tab:sw-chiral-cps-Zp}
\end{table}

We recall here the chiral couplings of the neutrinos, 
for a detailed discussion see Ref.~\cite{Iwamoto:2021wko}.
As the neutral currents are written in terms of flavor eigenstates,
the interactions between the neutral gauge bosons and the propagating
mass eigenstate neutrinos include also the neutrino mixing matrices:
\beq
\bom{\Gamma}^\mu_{V\nub_i\nu_j} = 
-\ri e \gamma^\mu
\Big(\bom{\Gamma}^L_{V\nub\nu} P_L + \bom{\Gamma}^R_{V\nub\nu} P_R\Big)_{ij}
\label{eq:GVnn}
\eeq
where
\beq
\bom{\Gamma}^L_{V\nub\nu} = 
 C^L_{V\nu\nu}\textbf{U}_L^\dagger \textbf{U}_L
-C^R_{V\nu\nu}\textbf{U}_R^T\textbf{U}_R^*
\label{eq:GLVnn}
\eeq
and 
\beq\label{eq:GRVnn}
\bom{\Gamma}^R_{V\nub\nu} = 
-C^L_{V\nu\nu}\textbf{U}_L^T \textbf{U}^*_L
+C^R_{V\nu\nu}\textbf{U}_R^\dagger\textbf{U}_R =
- \Big(\bom{\Gamma}^L_{V\nub\nu}\Big)^*
\eeq
for both $V=Z$ and $V=Z'$. 
In order to recover the SM vector and axial vector couplings of the $Z$ boson and the neutrinos, the right handed mixing matrices have to vanish and 
\be 
\textbf{U}_L^\dagger \textbf{U}_L \to \mathbf{1}_3
\quad\text{and}\quad
\textbf{U}_L^T \textbf{U}^*_L \to 0.
\ee
If one estimates the chiral couplings of the $Z$ and $Z'$ bosons in the presence of sterile neutrinos but with the mixing neglected then one needs to use the following replacements:
\be 
\bsp
\textbf{U}_L^\dagger \textbf{U}_L \to \mathbf{1}_3
\quad\text{and}&\quad
\textbf{U}_L^T \textbf{U}^*_L \to 0,
\\
\textbf{U}_R^\dagger \textbf{U}_R \to \mathbf{1}_3
\quad\text{and}&\quad
\textbf{U}_R^T \textbf{U}^*_R \to 0,
\esp
\ee
which we adopted throughout.

\section{Coefficient functions for hadroproduction and decays of the Z-prime boson}
\label{app:zp_decay}

The theoretical input needed to compute the Drell-Yan pair production cross section in Eq.~\eqref{eq:zp_prod1} is the coefficient functions and branching ratios, which we present here explicitly.
The coefficient functions for the production of a neutral gauge boson at NLO accuracy in QCD read \cite{Altarelli:1979ub}
\be 
\bsp
\Delta_{qq}\bigl(z, \mu_R^2\bigr) &= \delta(1-z) + 
\frac{\alpha_s(\mu_R^2)}{2\pi}
\\ & \times C_F\biggl[
\delta(1-z)\biggl(\frac{2\pi^2}{3}-8\biggr) + 4 \bigl(1+z^2\bigr) \biggl(\frac{\ln(1-z)}{1-z}\biggr)_{+}
- 2\frac{1+z^2}{1-z}\ln(z)
\biggr],
\\
\Delta_{gq}\bigl(z, \mu_R^2\bigr) &= \frac{\alpha_s(\mu_R^2)}{2\pi}\,
T_R\biggl[ (1-2z+2z^2)\ln\biggl(\frac{(1-z)^2}{z}\biggr)+\frac{1}{2}+3z - \frac{7}{2}z^2\biggr],
\esp
\ee
with color factors $C_F = 4/3$ and $T_R = 1/2$. 

The model considered here is defined in Sect.~\ref{sect:u1_extension}. The SM particle spectrum  is extended with the $Z'$ boson, a new scalar $s$ and 
three right handed neutrinos $N_i,~~i=1,2,3$. 
The mass of the new scalar $M_s$ has to be larger than half the mass of the Higgs boson $M_h$, otherwise the decay width of the Higgs particle becomes too large
as compared to the experimental upper limit $3.2^{+2.8}_{-2.2}\,\MeV$ \cite{CMS:2019ekd}. This means, that a light $Z'$ $\bigl(M_{Z'} < M_Z\bigr)$ can only decay into fermion pairs:
\be 
\Gamma(Z' \to f + \overline{f}) = 
N_C\, \rho \,C_{ff}\,M_{Z'}~\bigl(v_{Z',f}^2 + a_{Z',f}^2 \bigr) 
\label{eq:dec_zpZff}
\ee
where $\rho$ is defined in Eq.~\eqref{eq:u1_rho}, $C_{ff} = \frac{G_F M_Z^2}{6\sqrt{2}\pi}\simeq 3.6383\cdot 10^{-3}$, 
and the vector and axial vector couplings are given in Sect.~\ref{sect:VA-cps}.
Eq.~\eqref{eq:dec_zpZff} is valid for Dirac fermions. The formula for Majorana 
neutrinos can be obtained by the replacement $C_{ff} \to \frac12 C_{ff}$.
The invisible branching fraction of a light $Z'$ boson is 
\be 
\label{eq:inv_width}
\mathrm{Br}\bigl(Z' \to \text{inv.} \bigr) = \frac{3 n_N}{3 n_N + 3 (1- 2\cw^2 \mathscr{Z})^2 \, n_\ell +(1+2\cw^2 \mathscr{Z})^2 \, n_d +(1-4\cw^2 \mathscr{Z})^2 \, n_u},
\ee
where $n_f$ counts the kinematically allowed decays into $n_f$ families of 
fermion type $f$. The number $n_N$ counts the Dirac type neutrinos. 
The same formula applies for Majorana neutrinos with the replacement $n_N \to n_N/2$.
The parameter $\mathscr{Z}$ is defined in Eq.~\eqref{eq:x-def} and $\cw = \cos\theta_W$.

The larger $M_{Z'}$, the more decay channels are allowed. For the case of a heavy $Z'$, $M_{Z'} \gg M_h$, we neglect the finite mass effects of the particles $Z,W$ and $h$ in the following decay formulas. However, we keep the full dependence on the unknown $M_s$ as it is a free parameter of the model. 
The decays into a pair of charged $W$ bosons are \cite{Langacker:2008yv}
\be 
\bsp
\Gamma\bigr(Z' \to W^+ + W^-\bigl) &= \varepsilon M_{Z'}\frac{\xi^2 C_{ff}}{4 \rho}\,,
\\
\Gamma\bigr(Z' \to Z + W^+ + W^-\bigl) &= 
\varepsilon M_{Z'}\frac{7 \xi^4 C_{ff}^2}{320\pi}  \cw^4\,,
\\
\Gamma\bigr(Z' \to \gamma + W^+ + W^-\bigl) &= \varepsilon M_{Z'}\frac{301 \xi^2 C_{ff}^2 }{800\pi} \cw^2 \sw^2\,,
\esp
\ee
where $\varepsilon = (\xi \sz)^2$.
The decays into scalar particles also include two- and three-body ones as
\be 
\bsp
\Gamma\bigr(Z' \to Z + S\bigl) &= 
\varepsilon M_{Z'} \frac{\xi^2 C_{ff}}{4 \rho^2} 
\bigl|
\Gamma_{Z,Z',S_i}\bigr|^2
\,\bigl(1-\zeta_S^2\bigr)^3\,,
\\
\Gamma\bigr(Z' \to Z + S + S\bigl) &= \varepsilon M_{Z'} \frac{3 \xi^4 C_{ff}^2}{64 \pi \rho^2}
\bigl|\Gamma_{Z,Z',S_i,S_i}\bigr|^2
\biggl[
\sqrt{1-4\zeta_S^2}
\biggl(
1+\frac{26}{3}\zeta_{S}^2
-\frac{62}{3}
\zeta_{S}^4
+ 20\zeta_{S}^6
\biggr)
\\& 
- 4 \zeta_{S}^2
\bigl(1-
3\zeta_{S}^2 
+ 6\zeta_{S}^4  
- 5 \zeta_{S}^6
\bigr)
\\&\qquad\times
\ln\biggl(\frac{\bigl(1-2\zeta_S^2\bigr)\bigl(\sqrt{1-4\zeta_S^2}+1-2\zeta_S^2\bigr)}{2\zeta_S^4}-1\biggr)
\biggr],
\\
\Gamma\bigr(Z' \to Z + s + h\bigl) &= 
\varepsilon M_{Z'} \frac{3 \xi^4 C_{ff}^2 }{32 \pi \rho^2}
\biggl(\sss\cs +\frac{\sss\cs}{\tanb^2} \biggr)^2 
\\&\quad\times 
\biggl[1 +\frac{1}{3}\zeta_{S}^2\biggl(10 +
12\ln\bigl(\zeta_{S}^2\bigr) -18\zeta_{S}^2+6\zeta_{S}^4 -
\zeta_S^6\biggr)  \biggr],
\esp
\ee
where $S = h,s$, $\zeta_S = M_{S}/M_{Z'}$. For a heavy $Z'$, $M_{Z'} \gtrsim 1$\,TeV, we neglected $\zeta_h \lesssim \mathcal{O}\big(10^{-1}\big)$. The triple and quartic vector-scalar vertices are
\be 
\bsp
\Gamma_{Z,Z',h} = \cs + \sss/\tanb,
&\quad
\Gamma_{Z,Z',s} = \sss - \cs/\tanb,
\\
\Gamma_{Z,Z',h,h} = \cs^2 - \big(\sss/\tanb\big)^2,
&\quad
\Gamma_{Z,Z',s,s} = \sss^2 - \big(\cs/\tanb\big)^2.
\esp
\ee
The largest contribution from the scalar sector to the $Z'$ decay width is obtained by setting $\zeta_S =0$. In that case, 
we find the sum of partial decay widths with a scalar in the final state independent of the scalar mixing angle as
\be 
\label{eq:zp_to_scalar}
\sum \Gamma\bigl(Z' \to Z+ \text{ scalar} \bigr) \lesssim 
\varepsilon M_{Z'} \frac{1}{4\rho^2} \biggl[\xi^2 
C_{ff}\bigl(1+\tan^{-2}\beta\bigr)
+ \frac{3}{16\pi}
\xi^4 C_{ff}^2 
\bigl(1+\tan^{-4}\beta\bigr)\biggr].
\ee
We use this upper limit \eqref{eq:zp_to_scalar} in our numerical calculation to take into account the effect of the scalar sector in the total decay width of $Z'$.
Consulting Eq.~\eqref{eq:u1_rho} one can recognize that $\varepsilon  \simeq \rho-1$
is a small parameter for a heavy $Z'$. 

We computed the fermionic branching fractions of a heavy $Z'$ boson using the V-A couplings obtained in Sect.~\ref{sect:VA-cps}, and obtained 
\be 
\bsp
\mathrm{Br}\bigl(Z' \to \ell^+ \ell^- \bigr) &=  \frac{2 - 6 \mathscr{Z}  + 5 \mathscr{Z}^2}
{16 - 32 \mathscr{Z}  +  \mathscr{Z}^2\bigl( 41 + C_{w,s}\xi^2\bigr)},
\\
\mathrm{Br}\bigl(Z' \to U \overline{U} \bigr) &=  \frac{2 -10 \mathscr{Z}  + 17 \mathscr{Z}^2}
{48 - 96 \mathscr{Z}  +  3\mathscr{Z}^2\bigl( 41 + C_{w,s}\xi^2\bigr)},
\\
\mathrm{Br}\bigl(Z' \to D \overline{D} \bigr) &=  \frac{2 + 2 \mathscr{Z}  + 5 \mathscr{Z}^2}
{48 - 96 \mathscr{Z}  +  3\mathscr{Z}^2\bigl( 41 + C_{w,s}\xi^2\bigr)},
\esp
\ee
where the coupling constant is
\be 
C_{w,s} =  C_{ff}\frac{15+7\cw^4}{160\pi} \simeq 1.4 \times 10^{-4}
\,.
\ee
We note that for $M_{Z'} > 14~\TeV$, the decays into scalars and $W$ bosons 
start to dominate and the fermionic branching fractions may decrease significantly
depending on the charge assignment encoded in $\mathscr{Z}$.

\section{Kinetic mixing for U(1) extensions}
\label{app:na64_match}

Low energy experiments, such as NA64, BaBar, FASER place a stringent constraints on models which are extended by a $U(1)$ gauge group introducing a new gauge boson coupled to the SM fermions, which can be interpreted as a dark photon $A'$ that has kinetic mixing with the known photon. In the dark photon model the interaction term involving the $A'$ coupled to the electromagnetic current $J^\mu_{\rm EM}$ can be written as
\be 
\cL_{\rm int} = -\epsilon e A^{\prime}_\mu J^\mu_{\rm EM}
\ee
where $\epsilon$ can be viewed as the kinetic mixing parameter.
The experimental exclusion bounds are placed on the parameter plane $\bigl(\epsilon , M_{A'}\bigr)$. 
In order to extend those constraints to the parameters of a more general 
U(1) extension, one may set $M_{A'} \equiv M_{Z'}$ but 
relating $\epsilon$ to the free parameters discussed 
Sect.~\ref{sect:param-scan}.
involves some subtlety, discussed below.

The NA64 and BaBar experiments search for dark photon brehmsstrahlung 
in the invisible decay channel of the $A'$ \cite{NA64:2019imj}.
Following ref.~\cite{Iwamoto:2021fup} (see also \cite{Ilten:2018crw}) we have
\be 
\frac{\sigma\bigl(e^- +Z \longrightarrow e^- + Z + A' \bigr)}{\sigma\bigl(e^- +Z \longrightarrow e^- + Z + Z' \bigr)}=
\frac{\mathrm{Br}(Z' \to \mathrm{inv.})}{\mathrm{Br}(A' \to \mathrm{inv.})},
\label{eq:NA64matching}
\ee
where $\mathrm{Br}(Z' \to \mathrm{inv.})$
is given in Eq.~\eqref{eq:inv_width} and
and $\mathrm{Br}(A' \to \mathrm{inv.})=1$ in the 
NA64 experiment. 
Computing the cross sections of the $A'$ and $Z'$ bremsstrahlung processes yield
\be
\bsp
&
\frac{\sigma\bigl(e+Z \longrightarrow e + Z + A' \bigr)}{\sigma\bigl(e+Z \longrightarrow e + Z + Z' \bigr)}
=
\epsilon^2 \, \frac{\bigl(4 \sw^2 \cw^2 \bigr)}{ v_{Z',\ell}^2+ a_{Z',\ell}^2\bigl(1+f(M_{Z'})\bigr)} + \mathcal{O}\biggl(\frac{m_e^2}{s}\biggr),
\label{eq:xsectratio}
\esp
\ee
where $v_{Z',\ell}$ and $a_{Z',\ell}$ are given in Tab.~\ref{tab:sw-cps}, 
$\sqrt{s} = 100~\GeV$ in the NA64 experiment and 
\be 
f(M) = 2 \biggl(\frac{m_e^2}{M^2}\biggr)\biggl(1-\log^{-1}\biggl(\frac{M^2}{4s}\biggr) \biggr)+\mathcal{O}\biggl(\frac{m_e^4}{M^4}\biggr).
\ee
collects the finite mass effects of the electron and the $Z'$ boson.
Assembling the pieces for Eq.~\eqref{eq:NA64matching}, gives the matching relation 
\be 
\epsilon = \frac{\sqrt{ v_{Z',\ell}^2+ a_{Z',\ell}^2\bigl(1+f(M_{Z'})\bigr)}}{2\sw \cw}\sqrt{\mathrm{Br}(Z' \to \mathrm{inv.})}.
\label{eq:na64_swsm_epsilon}
\ee
The axial vector couplings can be neglected for a light 
$Z'$ boson, hence the matching relation reduces to
\be
\label{eq:match-eps-sz}
\epsilon = \frac{|\sz|}{2\sw \cw}\left|2 \cw^2 -\frac{1}{\mathscr{Z}}\right|
\sqrt{\mathrm{Br}(Z' \to \mathrm{inv.})}
\ee
or in terms of $g_z$ it is given as 
\be
\label{eq:match-eps-gz}
\epsilon = \frac{|z_N\,g_z |}{e}\left|2 \cw^2 \mathscr{Z} -1\right|
\sqrt{\mathrm{Br}(Z' \to \mathrm{inv.})}
\ee
where $\mathrm{Br}(Z' \to \mathrm{inv.})$ is given in \eqref{eq:inv_width}.
This formula reproduces the corresponding one given in Ref.~\cite{Iwamoto:2021fup} 
for $M_{Z'} < m_\pi = 130\,\MeV$ and $\mathscr{Z}=2$ (with $z_N = 1/2$ and $z_\phi = 1$), 
which is the superweak extension of the SM.

In the FASER experiment, predominantly a light neutral meson ($m^0 = \pi^0$, $\eta^0$ or $\omega^0$) decays into a neutral gauge boson pair, which may include the $m^ 0 \rightarrow \gamma + A'$ production channel. They searched for a dark photon in the $\gamma + A' \rightarrow \gamma + e^- + e^+$ decay channel \cite{FASER:2023tle}.
The partial rate of the neutral pion into a photon and dark photon is then given as 
\be 
\Gamma\bigl(\pi^0 \to A' + \gamma \bigr) 
= 2 \biggl(1 - \frac{M_{A'}^2}{m_\pi^2} \biggr)^3 
~\bigl( 2 ~\mathrm{tr}(\text{gens.}) ~\bigr)^2
~\Gamma\bigl(\pi^0 \to \gamma + \gamma \bigr),
\ee
where the first factor of $2$ is due to the symmetry factor difference between the  $ A' + \gamma$ and $\gamma + \gamma$ final states and the factor containing $M_{A'}$ is due to the polarization sum of a massive vector boson times the phase space factor. The trace of the generators is 
\be 
2 ~\mathrm{tr}(\text{gens.}) = 2  N_C\mathrm{tr}\bigl(\tau^a Q \, Q\bigr) ~ \epsilon = \epsilon,
\ee
with the matrices
\be 
\tau^a = \begin{pmatrix}
  1/2 & 0\\ 
  0 & -1/2
\end{pmatrix},
\quad\text{and}\quad
Q = \begin{pmatrix}
  2/3 & 0\\ 
  0 & -1/3
\end{pmatrix}.
\ee
The trace of the generators in $U(1)$ extensions considered in this paper is 
\be 
\label{eq:FASER-match}
2 ~\mathrm{tr}(\text{gens.}) = 2 N_C\mathrm{tr}\bigl(\tau^a Q\, \mathbf{v}_{Z',q}\bigr) = 
\biggl|\frac{v_{Z',\ell}}{2\sw \cw}\biggr|,
\ee
where 
\be 
\mathbf{v}_{Z',q} = 
\begin{pmatrix}
  v_{Z',u} & 0\\ 
  0 & v_{Z',d}
\end{pmatrix}
= 
\frac{\sz}{3}
\begin{pmatrix}
  \frac{1}{\mathscr{Z}}-4\cw^2 & 0\\ 
  0 & \frac{1}{\mathscr{Z}}+2\cw^2
\end{pmatrix}.
\ee
In order to match the exclusion bounds of FASER one has to solve two equations for the two parameters $(M_{A'},\epsilon)$. The first one expresses the equality of the signal events in the two models:
\be
\Gamma\bigl(\pi^0 \to A' + \gamma \bigr)\,\mathrm{Br}(A'\to e^+ e^-)
= 
\Gamma\bigl(\pi^0 \to Z' + \gamma \bigr)\,\mathrm{Br}(Z' \to e^+ e^-)
\ee
where $\mathrm{Br}(A' \to e^+ e^-) = 1$ in the FASER experiment. The second equation,
\be 
M_{A'}\Gamma_{A'} = M_{Z'}\Gamma_{Z'}\,,
\ee
ensures that the decay length of the dark photon $A'$ and that of 
the $Z'$ boson are the same as those are required to decay in the 
detector itself. For $M_{A'}\ll m_\pi$ the matching relations yield
\be
\bsp
M_{Z'} &= \mathrm{Br}(Z' \to e^+ e^-) M_{A'},
\\
|z_N\, g_z| &= \frac{e \epsilon}{\sqrt{\mathrm{Br}(Z' \to e^+ e^-)}} \frac{1}{|1-2\cw^2 \mathscr{Z}|}.
\esp
\ee
In the $B-L$ extension considered in Ref.~\cite{FASER:2023tle} one has $\mathscr{Z} = 0$, $\mathrm{Br}(Z' \to e^+ e^-) = 2/5$ and 
$z_N = 1$ as they use a different normalization for the $z$ charges. 


%

\begin{thebibliography}{37}%
\makeatletter
\providecommand \@ifxundefined [1]{%
 \@ifx{#1\undefined}
}%
\providecommand \@ifnum [1]{%
 \ifnum #1\expandafter \@firstoftwo
 \else \expandafter \@secondoftwo
 \fi
}%
\providecommand \@ifx [1]{%
 \ifx #1\expandafter \@firstoftwo
 \else \expandafter \@secondoftwo
 \fi
}%
\providecommand \natexlab [1]{#1}%
\providecommand \enquote  [1]{``#1''}%
\providecommand \bibnamefont  [1]{#1}%
\providecommand \bibfnamefont [1]{#1}%
\providecommand \citenamefont [1]{#1}%
\providecommand \href@noop [0]{\@secondoftwo}%
\providecommand \href [0]{\begingroup \@sanitize@url \@href}%
\providecommand \@href[1]{\@@startlink{#1}\@@href}%
\providecommand \@@href[1]{\endgroup#1\@@endlink}%
\providecommand \@sanitize@url [0]{\catcode `\\12\catcode `\$12\catcode
  `\&12\catcode `\#12\catcode `\^12\catcode `\_12\catcode `\%12\relax}%
\providecommand \@@startlink[1]{}%
\providecommand \@@endlink[0]{}%
\providecommand \url  [0]{\begingroup\@sanitize@url \@url }%
\providecommand \@url [1]{\endgroup\@href {#1}{\urlprefix }}%
\providecommand \urlprefix  [0]{URL }%
\providecommand \Eprint [0]{\href }%
\providecommand \doibase [0]{http://dx.doi.org/}%
\providecommand \selectlanguage [0]{\@gobble}%
\providecommand \bibinfo  [0]{\@secondoftwo}%
\providecommand \bibfield  [0]{\@secondoftwo}%
\providecommand \translation [1]{[#1]}%
\providecommand \BibitemOpen [0]{}%
\providecommand \bibitemStop [0]{}%
\providecommand \bibitemNoStop [0]{.\EOS\space}%
\providecommand \EOS [0]{\spacefactor3000\relax}%
\providecommand \BibitemShut  [1]{\csname bibitem#1\endcsname}%
\let\auto@bib@innerbib\@empty
\bibitem [{\citenamefont {Workman}\ and\ \citenamefont
  {Others}(2022)}]{Workman:2022ynf}%
  \BibitemOpen
  \bibfield  {author} {\bibinfo {author} {\bibfnamefont {R.~L.}\ \bibnamefont
  {Workman}}\ and\ \bibinfo {author} {\bibnamefont {Others}} (\bibinfo
  {collaboration} {Particle Data Group}),\ }\href {\doibase
  10.1093/ptep/ptac097} {\bibfield  {journal} {\bibinfo  {journal} {PTEP}\
  }\textbf {\bibinfo {volume} {2022}},\ \bibinfo {pages} {083C01} (\bibinfo
  {year} {2022})}\BibitemShut {NoStop}%
\bibitem{ATLAS:2022djm}
 G.~Aad et al.~[ATLAS],
\href{https://cds.cern.ch/record/2804061}{ATL-PHYS-PUB-2022-009}.

\bibitem{CMS-SM}
\href{https://cms-results.web.cern.ch/cms-results/public-results/publications/SMP}{https://cms-results.web.cern.ch/cms-results/public-results/publications/SMP}%

\bibitem [{\citenamefont {Aguillard}\ \emph {et~al.}(2023)\citenamefont
  {Aguillard} \emph {et~al.}}]{Muong-2:2023cdq}%
  \BibitemOpen
  \bibfield  {author} {\bibinfo {author} {\bibfnamefont {D.~P.}\ \bibnamefont
  {Aguillard}} \emph {et~al.} (\bibinfo {collaboration} {Muon g-2}),\ }\href
  {\doibase 10.1103/PhysRevLett.131.161802} {\bibfield  {journal} {\bibinfo
  {journal} {Phys. Rev. Lett.}\ }\textbf {\bibinfo {volume} {131}},\ \bibinfo
  {pages} {161802} (\bibinfo {year} {2023})},\ \Eprint
  {http://arxiv.org/abs/2308.06230} {arXiv:2308.06230 [hep-ex]} \BibitemShut
  {NoStop}%
\bibitem [{\citenamefont {Aoyama}\ \emph {et~al.}(2020)\citenamefont {Aoyama}
  \emph {et~al.}}]{Aoyama:2020ynm}%
  \BibitemOpen
  \bibfield  {author} {\bibinfo {author} {\bibfnamefont {T.}~\bibnamefont
  {Aoyama}} \emph {et~al.},\ }\href {\doibase 10.1016/j.physrep.2020.07.006}
  {\bibfield  {journal} {\bibinfo  {journal} {Phys. Rept.}\ }\textbf {\bibinfo
  {volume} {887}},\ \bibinfo {pages} {1} (\bibinfo {year} {2020})},\ \Eprint
  {http://arxiv.org/abs/2006.04822} {arXiv:2006.04822 [hep-ph]} \BibitemShut
  {NoStop}%
\bibitem [{\citenamefont {Pospelov}(2009)}]{Pospelov:2008zw}%
  \BibitemOpen
  \bibfield  {author} {\bibinfo {author} {\bibfnamefont {M.}~\bibnamefont
  {Pospelov}},\ }\href {\doibase 10.1103/PhysRevD.80.095002} {\bibfield
  {journal} {\bibinfo  {journal} {Phys. Rev. D}\ }\textbf {\bibinfo {volume}
  {80}},\ \bibinfo {pages} {095002} (\bibinfo {year} {2009})},\ \Eprint
  {http://arxiv.org/abs/0811.1030} {arXiv:0811.1030 [hep-ph]} \BibitemShut
  {NoStop}%
\bibitem [{\citenamefont {Davoudiasl}\ \emph {et~al.}(2012)\citenamefont
  {Davoudiasl}, \citenamefont {Lee},\ and\ \citenamefont
  {Marciano}}]{Davoudiasl:2012ig}%
  \BibitemOpen
  \bibfield  {author} {\bibinfo {author} {\bibfnamefont {H.}~\bibnamefont
  {Davoudiasl}}, \bibinfo {author} {\bibfnamefont {H.-S.}\ \bibnamefont {Lee}},
  \ and\ \bibinfo {author} {\bibfnamefont {W.~J.}\ \bibnamefont {Marciano}},\
  }\href {\doibase 10.1103/PhysRevD.86.095009} {\bibfield  {journal} {\bibinfo
  {journal} {Phys. Rev. D}\ }\textbf {\bibinfo {volume} {86}},\ \bibinfo
  {pages} {095009} (\bibinfo {year} {2012})},\ \Eprint
  {http://arxiv.org/abs/1208.2973} {arXiv:1208.2973 [hep-ph]} \BibitemShut
  {NoStop}%
\bibitem [{\citenamefont {Cadeddu}\ \emph {et~al.}(2021)\citenamefont
  {Cadeddu}, \citenamefont {Cargioli}, \citenamefont {Dordei}, \citenamefont
  {Giunti},\ and\ \citenamefont {Picciau}}]{Cadeddu:2021dqx}%
  \BibitemOpen
  \bibfield  {author} {\bibinfo {author} {\bibfnamefont {M.}~\bibnamefont
  {Cadeddu}}, \bibinfo {author} {\bibfnamefont {N.}~\bibnamefont {Cargioli}},
  \bibinfo {author} {\bibfnamefont {F.}~\bibnamefont {Dordei}}, \bibinfo
  {author} {\bibfnamefont {C.}~\bibnamefont {Giunti}}, \ and\ \bibinfo {author}
  {\bibfnamefont {E.}~\bibnamefont {Picciau}},\ }\href {\doibase
  10.1103/PhysRevD.104.L011701} {\bibfield  {journal} {\bibinfo  {journal}
  {Phys. Rev. D}\ }\textbf {\bibinfo {volume} {104}},\ \bibinfo {pages}
  {011701} (\bibinfo {year} {2021})},\ \Eprint
  {http://arxiv.org/abs/2104.03280} {arXiv:2104.03280 [hep-ph]} \BibitemShut
  {NoStop}%
\bibitem [{\citenamefont {Borsanyi}\ \emph {et~al.}(2021)\citenamefont
  {Borsanyi} \emph {et~al.}}]{Borsanyi:2020mff}%
  \BibitemOpen
  \bibfield  {author} {\bibinfo {author} {\bibfnamefont {S.}~\bibnamefont
  {Borsanyi}} \emph {et~al.},\ }\href {\doibase 10.1038/s41586-021-03418-1}
  {\bibfield  {journal} {\bibinfo  {journal} {Nature}\ }\textbf {\bibinfo
  {volume} {593}},\ \bibinfo {pages} {51} (\bibinfo {year} {2021})},\ \Eprint
  {http://arxiv.org/abs/2002.12347} {arXiv:2002.12347 [hep-lat]} \BibitemShut
  {NoStop}%
\bibitem [{\citenamefont {Okun}(1982)}]{Okun:1982xi}%
  \BibitemOpen
  \bibfield  {author} {\bibinfo {author} {\bibfnamefont {L.~B.}\ \bibnamefont
  {Okun}},\ }\href@noop {} {\bibfield  {journal} {\bibinfo  {journal} {Sov.
  Phys. JETP}\ }\textbf {\bibinfo {volume} {56}},\ \bibinfo {pages} {502}
  (\bibinfo {year} {1982})}\BibitemShut {NoStop}%
\bibitem [{\citenamefont {He}\ \emph {et~al.}(1991)\citenamefont {He},
  \citenamefont {Joshi}, \citenamefont {Lew},\ and\ \citenamefont
  {Volkas}}]{He:1990pn}%
  \BibitemOpen
  \bibfield  {author} {\bibinfo {author} {\bibfnamefont {X.~G.}\ \bibnamefont
  {He}}, \bibinfo {author} {\bibfnamefont {G.~C.}\ \bibnamefont {Joshi}},
  \bibinfo {author} {\bibfnamefont {H.}~\bibnamefont {Lew}}, \ and\ \bibinfo
  {author} {\bibfnamefont {R.~R.}\ \bibnamefont {Volkas}},\ }\href {\doibase
  10.1103/PhysRevD.43.R22} {\bibfield  {journal} {\bibinfo  {journal} {Phys.
  Rev. D}\ }\textbf {\bibinfo {volume} {43}},\ \bibinfo {pages} {22} (\bibinfo
  {year} {1991})}\BibitemShut {NoStop}%
\bibitem [{\citenamefont {Gopalakrishna}\ \emph {et~al.}(2008)\citenamefont
  {Gopalakrishna}, \citenamefont {Jung},\ and\ \citenamefont
  {Wells}}]{Gopalakrishna:2008dv}%
  \BibitemOpen
  \bibfield  {author} {\bibinfo {author} {\bibfnamefont {S.}~\bibnamefont
  {Gopalakrishna}}, \bibinfo {author} {\bibfnamefont {S.}~\bibnamefont {Jung}},
  \ and\ \bibinfo {author} {\bibfnamefont {J.~D.}\ \bibnamefont {Wells}},\
  }\href {\doibase 10.1103/PhysRevD.78.055002} {\bibfield  {journal} {\bibinfo
  {journal} {Phys. Rev. D}\ }\textbf {\bibinfo {volume} {78}},\ \bibinfo
  {pages} {055002} (\bibinfo {year} {2008})},\ \Eprint
  {http://arxiv.org/abs/0801.3456} {arXiv:0801.3456 [hep-ph]} \BibitemShut
  {NoStop}%
\bibitem [{Note1()}]{Note1}%
  \BibitemOpen
  \bibinfo {note} {After diagonalization of the mass matrix of the neutral
  gauge bosons}\BibitemShut {NoStop}%
\bibitem [{\citenamefont {Tr\'ocs\'anyi}(2020)}]{Trocsanyi:2018bkm}%
  \BibitemOpen
  \bibfield  {author} {\bibinfo {author} {\bibfnamefont {Z.}~\bibnamefont
  {Tr\'ocs\'anyi}},\ }\href {\doibase 10.3390/sym12010107} {\bibfield
  {journal} {\bibinfo  {journal} {Symmetry}\ }\textbf {\bibinfo {volume}
  {12}},\ \bibinfo {pages} {107} (\bibinfo {year} {2020})},\ \Eprint
  {http://arxiv.org/abs/1812.11189} {arXiv:1812.11189 [hep-ph]} \BibitemShut
  {NoStop}%
\bibitem [{\citenamefont {Dittmaier}\ \emph {et~al.}(2023)\citenamefont
  {Dittmaier}, \citenamefont {Rehberg},\ and\ \citenamefont
  {Rzehak}}]{Dittmaier:2023ovi}%
  \BibitemOpen
  \bibfield  {author} {\bibinfo {author} {\bibfnamefont {S.}~\bibnamefont
  {Dittmaier}}, \bibinfo {author} {\bibfnamefont {J.}~\bibnamefont {Rehberg}},
  \ and\ \bibinfo {author} {\bibfnamefont {H.}~\bibnamefont {Rzehak}},\
  }\href@noop {} {\  (\bibinfo {year} {2023})},\ \Eprint
  {http://arxiv.org/abs/2308.07845} {arXiv:2308.07845 [hep-ph]} \BibitemShut
  {NoStop}%
\bibitem [{\citenamefont {Ilten}\ \emph {et~al.}(2018)\citenamefont {Ilten},
  \citenamefont {Soreq}, \citenamefont {Williams},\ and\ \citenamefont
  {Xue}}]{Ilten:2018crw}%
  \BibitemOpen
  \bibfield  {author} {\bibinfo {author} {\bibfnamefont {P.}~\bibnamefont
  {Ilten}}, \bibinfo {author} {\bibfnamefont {Y.}~\bibnamefont {Soreq}},
  \bibinfo {author} {\bibfnamefont {M.}~\bibnamefont {Williams}}, \ and\
  \bibinfo {author} {\bibfnamefont {W.}~\bibnamefont {Xue}},\ }\href {\doibase
  10.1007/JHEP06(2018)004} {\bibfield  {journal} {\bibinfo  {journal} {JHEP}\
  }\textbf {\bibinfo {volume} {06}},\ \bibinfo {pages} {004} (\bibinfo {year}
  {2018})},\ \Eprint {http://arxiv.org/abs/1801.04847} {arXiv:1801.04847
  [hep-ph]} \BibitemShut {NoStop}%
\bibitem [{\citenamefont {Abada}\ \emph {et~al.}(2019)\citenamefont {Abada}
  \emph {et~al.}}]{FCC:2018bvk}%
  \BibitemOpen
  \bibfield  {author} {\bibinfo {author} {\bibfnamefont {A.}~\bibnamefont
  {Abada}} \emph {et~al.} (\bibinfo {collaboration} {FCC}),\ }\href {\doibase
  10.1140/epjst/e2019-900088-6} {\bibfield  {journal} {\bibinfo  {journal}
  {Eur. Phys. J. ST}\ }\textbf {\bibinfo {volume} {228}},\ \bibinfo {pages}
  {1109} (\bibinfo {year} {2019})}\BibitemShut {NoStop}%
\bibitem [{\citenamefont {Feng}\ \emph {et~al.}(2018)\citenamefont {Feng},
  \citenamefont {Galon}, \citenamefont {Kling},\ and\ \citenamefont
  {Trojanowski}}]{Feng:2017uoz}%
  \BibitemOpen
  \bibfield  {author} {\bibinfo {author} {\bibfnamefont {J.~L.}\ \bibnamefont
  {Feng}}, \bibinfo {author} {\bibfnamefont {I.}~\bibnamefont {Galon}},
  \bibinfo {author} {\bibfnamefont {F.}~\bibnamefont {Kling}}, \ and\ \bibinfo
  {author} {\bibfnamefont {S.}~\bibnamefont {Trojanowski}},\ }\href {\doibase
  10.1103/PhysRevD.97.035001} {\bibfield  {journal} {\bibinfo  {journal} {Phys.
  Rev. D}\ }\textbf {\bibinfo {volume} {97}},\ \bibinfo {pages} {035001}
  (\bibinfo {year} {2018})},\ \Eprint {http://arxiv.org/abs/1708.09389}
  {arXiv:1708.09389 [hep-ph]} \BibitemShut {NoStop}%
\bibitem [{Note2()}]{Note2}%
  \BibitemOpen
  \bibinfo {note} {The cases of an intermediate mass $Z'$, $M_{Z'} \approx
  M_Z$, and a very light $Z'$, $M_{Z'} \lesssim 20$\protect \,MeV have been
  discussed elsewhere.}\BibitemShut {Stop}%
\bibitem [{\citenamefont {Aad}\ \emph {et~al.}(2019)\citenamefont {Aad} \emph
  {et~al.}}]{ATLAS:2019erb}%
  \BibitemOpen
  \bibfield  {author} {\bibinfo {author} {\bibfnamefont {G.}~\bibnamefont
  {Aad}} \emph {et~al.} (\bibinfo {collaboration} {ATLAS}),\ }\href {\doibase
  10.1016/j.physletb.2019.07.016} {\bibfield  {journal} {\bibinfo  {journal}
  {Phys. Lett. B}\ }\textbf {\bibinfo {volume} {796}},\ \bibinfo {pages} {68}
  (\bibinfo {year} {2019})},\ \Eprint {http://arxiv.org/abs/1903.06248}
  {arXiv:1903.06248 [hep-ex]} \BibitemShut {NoStop}%
\bibitem [{\citenamefont {Sirunyan}\ \emph {et~al.}(2021)\citenamefont
  {Sirunyan} \emph {et~al.}}]{CMS:2021ctt}%
  \BibitemOpen
  \bibfield  {author} {\bibinfo {author} {\bibfnamefont {A.~M.}\ \bibnamefont
  {Sirunyan}} \emph {et~al.} (\bibinfo {collaboration} {CMS}),\ }\href
  {\doibase 10.1007/JHEP07(2021)208} {\bibfield  {journal} {\bibinfo  {journal}
  {JHEP}\ }\textbf {\bibinfo {volume} {07}},\ \bibinfo {pages} {208} (\bibinfo
  {year} {2021})},\ \Eprint {http://arxiv.org/abs/2103.02708} {arXiv:2103.02708
  [hep-ex]} \BibitemShut {NoStop}%
\bibitem [{\citenamefont {Banerjee}\ \emph {et~al.}(2019)\citenamefont
  {Banerjee} \emph {et~al.}}]{NA64:2019imj}%
  \BibitemOpen
  \bibfield  {author} {\bibinfo {author} {\bibfnamefont {D.}~\bibnamefont
  {Banerjee}} \emph {et~al.} (\bibinfo {collaboration} {{NA64}}),\ }\href
  {\doibase 10.1103/PhysRevLett.123.121801} {\bibfield  {journal} {\bibinfo
  {journal} {Phys. Rev. Lett.}\ }\textbf {\bibinfo {volume} {123}},\ \bibinfo
  {pages} {121801} (\bibinfo {year} {2019})},\ \Eprint
  {http://arxiv.org/abs/1906.00176} {arXiv:1906.00176 [hep-ex]} \BibitemShut
  {NoStop}%
\bibitem [{\citenamefont {Lees}\ \emph {et~al.}(2017)\citenamefont {Lees} \emph
  {et~al.}}]{BaBar:2017tiz}%
  \BibitemOpen
  \bibfield  {author} {\bibinfo {author} {\bibfnamefont {J.~P.}\ \bibnamefont
  {Lees}} \emph {et~al.} (\bibinfo {collaboration} {BaBar}),\ }\href {\doibase
  10.1103/PhysRevLett.119.131804} {\bibfield  {journal} {\bibinfo  {journal}
  {Phys. Rev. Lett.}\ }\textbf {\bibinfo {volume} {119}},\ \bibinfo {pages}
  {131804} (\bibinfo {year} {2017})},\ \Eprint
  {http://arxiv.org/abs/1702.03327} {arXiv:1702.03327 [hep-ex]} \BibitemShut
  {NoStop}%
\bibitem [{\citenamefont {Abreu}\ \emph {et~al.}(2024)\citenamefont {Abreu}
  \emph {et~al.}}]{FASER:2023tle}%
  \BibitemOpen
  \bibfield  {author} {\bibinfo {author} {\bibfnamefont {H.}~\bibnamefont
  {Abreu}} \emph {et~al.} (\bibinfo {collaboration} {FASER}),\ }\href {\doibase
  10.1016/j.physletb.2023.138378} {\bibfield  {journal} {\bibinfo  {journal}
  {Phys. Lett. B}\ }\textbf {\bibinfo {volume} {848}},\ \bibinfo {pages}
  {138378} (\bibinfo {year} {2024})},\ \Eprint
  {http://arxiv.org/abs/2308.05587} {arXiv:2308.05587 [hep-ex]} \BibitemShut
  {NoStop}%
\bibitem [{\citenamefont {Asai}\ \emph {et~al.}(2022)\citenamefont {Asai},
  \citenamefont {Das}, \citenamefont {Li}, \citenamefont {Nomura},\ and\
  \citenamefont {Seto}}]{Asai:2022zxw}%
  \BibitemOpen
  \bibfield  {author} {\bibinfo {author} {\bibfnamefont {K.}~\bibnamefont
  {Asai}}, \bibinfo {author} {\bibfnamefont {A.}~\bibnamefont {Das}}, \bibinfo
  {author} {\bibfnamefont {J.}~\bibnamefont {Li}}, \bibinfo {author}
  {\bibfnamefont {T.}~\bibnamefont {Nomura}}, \ and\ \bibinfo {author}
  {\bibfnamefont {O.}~\bibnamefont {Seto}},\ }\href {\doibase
  10.1103/PhysRevD.106.095033} {\bibfield  {journal} {\bibinfo  {journal}
  {Phys. Rev. D}\ }\textbf {\bibinfo {volume} {106}},\ \bibinfo {pages}
  {095033} (\bibinfo {year} {2022})},\ \Eprint
  {http://arxiv.org/abs/2206.12676} {arXiv:2206.12676 [hep-ph]} \BibitemShut
  {NoStop}%
\bibitem [{\citenamefont {Asai}\ \emph {et~al.}(2023)\citenamefont {Asai},
  \citenamefont {Das}, \citenamefont {Li}, \citenamefont {Nomura},\ and\
  \citenamefont {Seto}}]{Asai:2023xxl}%
  \BibitemOpen
  \bibfield  {author} {\bibinfo {author} {\bibfnamefont {K.}~\bibnamefont
  {Asai}}, \bibinfo {author} {\bibfnamefont {A.}~\bibnamefont {Das}}, \bibinfo
  {author} {\bibfnamefont {J.}~\bibnamefont {Li}}, \bibinfo {author}
  {\bibfnamefont {T.}~\bibnamefont {Nomura}}, \ and\ \bibinfo {author}
  {\bibfnamefont {O.}~\bibnamefont {Seto}},\ }\href@noop {} {\  (\bibinfo
  {year} {2023})},\ \Eprint {http://arxiv.org/abs/2307.09737} {arXiv:2307.09737
  [hep-ph]} \BibitemShut {NoStop}%
\bibitem [{Note3()}]{Note3}%
  \BibitemOpen
  \bibinfo {note} {This is the convention used in \protect \texttt {SARAH} for
  models with multiple gauged U(1) symmetries.}\BibitemShut {Stop}%
\bibitem [{\citenamefont {Iwamoto}\ \emph {et~al.}(2022)\citenamefont
  {Iwamoto}, \citenamefont {Seller},\ and\ \citenamefont
  {Tr\'ocs\'anyi}}]{Iwamoto:2021fup}%
  \BibitemOpen
  \bibfield  {author} {\bibinfo {author} {\bibfnamefont {S.}~\bibnamefont
  {Iwamoto}}, \bibinfo {author} {\bibfnamefont {K.}~\bibnamefont {Seller}}, \
  and\ \bibinfo {author} {\bibfnamefont {Z.}~\bibnamefont {Tr\'ocs\'anyi}},\
  }\href {\doibase 10.1088/1475-7516/2022/01/035} {\bibfield  {journal}
  {\bibinfo  {journal} {JCAP}\ }\textbf {\bibinfo {volume} {01}},\ \bibinfo
  {pages} {035} (\bibinfo {year} {2022})},\ \Eprint
  {http://arxiv.org/abs/2104.11248} {arXiv:2104.11248 [hep-ph]} \BibitemShut
  {NoStop}%
\bibitem [{\citenamefont {Appelquist}\ \emph {et~al.}(2003)\citenamefont
  {Appelquist}, \citenamefont {Dobrescu},\ and\ \citenamefont
  {Hopper}}]{Appelquist:2002mw}%
  \BibitemOpen
  \bibfield  {author} {\bibinfo {author} {\bibfnamefont {T.}~\bibnamefont
  {Appelquist}}, \bibinfo {author} {\bibfnamefont {B.~A.}\ \bibnamefont
  {Dobrescu}}, \ and\ \bibinfo {author} {\bibfnamefont {A.~R.}\ \bibnamefont
  {Hopper}},\ }\href {\doibase 10.1103/PhysRevD.68.035012} {\bibfield
  {journal} {\bibinfo  {journal} {Phys. Rev. D}\ }\textbf {\bibinfo {volume}
  {68}},\ \bibinfo {pages} {035012} (\bibinfo {year} {2003})},\ \Eprint
  {http://arxiv.org/abs/hep-ph/0212073} {arXiv:hep-ph/0212073} \BibitemShut
  {NoStop}%
\bibitem [{Note4()}]{Note4}%
  \BibitemOpen
  \bibinfo {note} {Our sign convention for $\theta _Z$ agrees with the
  convention of Ref.~\cite {Trocsanyi:2018bkm}, which differs by a factor of
  $(-1)$ from that of Ref.~\cite {Iwamoto:2021fup}.}\BibitemShut {Stop}%
\bibitem [{\citenamefont {Carena}\ \emph {et~al.}(2004)\citenamefont {Carena},
  \citenamefont {Daleo}, \citenamefont {Dobrescu},\ and\ \citenamefont
  {Tait}}]{Carena:2004xs}%
  \BibitemOpen
  \bibfield  {author} {\bibinfo {author} {\bibfnamefont {M.}~\bibnamefont
  {Carena}}, \bibinfo {author} {\bibfnamefont {A.}~\bibnamefont {Daleo}},
  \bibinfo {author} {\bibfnamefont {B.~A.}\ \bibnamefont {Dobrescu}}, \ and\
  \bibinfo {author} {\bibfnamefont {T.~M.~P.}\ \bibnamefont {Tait}},\ }\href
  {\doibase 10.1103/PhysRevD.70.093009} {\bibfield  {journal} {\bibinfo
  {journal} {Phys. Rev. D}\ }\textbf {\bibinfo {volume} {70}},\ \bibinfo
  {pages} {093009} (\bibinfo {year} {2004})},\ \Eprint
  {http://arxiv.org/abs/hep-ph/0408098} {arXiv:hep-ph/0408098} \BibitemShut
  {NoStop}%
\bibitem [{\citenamefont {Altarelli}\ \emph {et~al.}(1979)\citenamefont
  {Altarelli}, \citenamefont {Ellis},\ and\ \citenamefont
  {Martinelli}}]{Altarelli:1979ub}%
  \BibitemOpen
  \bibfield  {author} {\bibinfo {author} {\bibfnamefont {G.}~\bibnamefont
  {Altarelli}}, \bibinfo {author} {\bibfnamefont {R.~K.}\ \bibnamefont
  {Ellis}}, \ and\ \bibinfo {author} {\bibfnamefont {G.}~\bibnamefont
  {Martinelli}},\ }\href {\doibase 10.1016/0550-3213(79)90116-0} {\bibfield
  {journal} {\bibinfo  {journal} {Nucl. Phys. B}\ }\textbf {\bibinfo {volume}
  {157}},\ \bibinfo {pages} {461} (\bibinfo {year} {1979})}\BibitemShut
  {NoStop}%
\bibitem [{Man(2017)}]{Mangano:2017tke}%
  \BibitemOpen
  \href {\doibase 10.23731/CYRM-2017-003} {\ \textbf {\bibinfo {volume}
  {3/2017}} (\bibinfo {year} {2017}),\ 10.23731/CYRM-2017-003},\ \Eprint
  {http://arxiv.org/abs/1710.06353} {arXiv:1710.06353 [hep-ph]} \BibitemShut
  {NoStop}%
\bibitem [{\citenamefont {P\'eli}\ and\ \citenamefont
  {Tr\'ocs\'anyi}(2023)}]{Peli:2023fyb}%
  \BibitemOpen
  \bibfield  {author} {\bibinfo {author} {\bibfnamefont {Z.}~\bibnamefont
  {P\'eli}}\ and\ \bibinfo {author} {\bibfnamefont {Z.}~\bibnamefont
  {Tr\'ocs\'anyi}},\ }\href {\doibase 10.1103/PhysRevD.108.L031704} {\bibfield
  {journal} {\bibinfo  {journal} {Phys. Rev. D}\ }\textbf {\bibinfo {volume}
  {108}},\ \bibinfo {pages} {L031704} (\bibinfo {year} {2023})},\ \Eprint
  {http://arxiv.org/abs/2305.11931} {arXiv:2305.11931 [hep-ph]} \BibitemShut
  {NoStop}%
\bibitem [{\citenamefont {Athron}\ \emph {et~al.}(2022)\citenamefont {Athron},
  \citenamefont {Bach}, \citenamefont {Jacob}, \citenamefont {Kotlarski},
  \citenamefont {St\"ockinger},\ and\ \citenamefont {Voigt}}]{Athron:2022isz}%
  \BibitemOpen
  \bibfield  {author} {\bibinfo {author} {\bibfnamefont {P.}~\bibnamefont
  {Athron}}, \bibinfo {author} {\bibfnamefont {M.}~\bibnamefont {Bach}},
  \bibinfo {author} {\bibfnamefont {D.~H.~J.}\ \bibnamefont {Jacob}}, \bibinfo
  {author} {\bibfnamefont {W.}~\bibnamefont {Kotlarski}}, \bibinfo {author}
  {\bibfnamefont {D.}~\bibnamefont {St\"ockinger}}, \ and\ \bibinfo {author}
  {\bibfnamefont {A.}~\bibnamefont {Voigt}},\ }\href {\doibase
  10.1103/PhysRevD.106.095023} {\bibfield  {journal} {\bibinfo  {journal}
  {Phys. Rev. D}\ }\textbf {\bibinfo {volume} {106}},\ \bibinfo {pages}
  {095023} (\bibinfo {year} {2022})},\ \Eprint
  {http://arxiv.org/abs/2204.05285} {arXiv:2204.05285 [hep-ph]} \BibitemShut
  {NoStop}%
\bibitem [{\citenamefont {P\'eli}(2023)}]{Peli:2023fxw}%
  \BibitemOpen
  \bibfield  {author} {\bibinfo {author} {\bibfnamefont {Z.}~\bibnamefont
  {P\'eli}},\ }in\ \href@noop {} {\emph {\bibinfo {booktitle} {{45th
  International Conference of Theoretical Physics}: {Matter To The Deepest
  Recent Developments In Physics Of Fundamental Interactions}}}}\ (\bibinfo
  {year} {2023})\ \Eprint {http://arxiv.org/abs/2311.08203} {arXiv:2311.08203
  [hep-ph]} \BibitemShut {NoStop}%
\bibitem{CERN-ACC-2019-028}  
C.~Helsens, D.~Jamin, M.~Selvaggi,
\Eprint {https://cds.cern.ch/record/2642473/files/CERN-ACC-2019-0028.pdf} {CERN-ACC-2019-028}.%
\bibitem [{\citenamefont {Iwamoto}\ \emph {et~al.}(2021)\citenamefont
  {Iwamoto}, \citenamefont {K\"arkk\"ainen}, \citenamefont {P\'eli},\ and\
  \citenamefont {Tr\'ocs\'anyi}}]{Iwamoto:2021wko}%
  \BibitemOpen
  \bibfield  {author} {\bibinfo {author} {\bibfnamefont {S.}~\bibnamefont
  {Iwamoto}}, \bibinfo {author} {\bibfnamefont {T.~J.}\ \bibnamefont
  {K\"arkk\"ainen}}, \bibinfo {author} {\bibfnamefont {Z.}~\bibnamefont
  {P\'eli}}, \ and\ \bibinfo {author} {\bibfnamefont {Z.}~\bibnamefont
  {Tr\'ocs\'anyi}},\ }\href {\doibase 10.1103/PhysRevD.104.055042} {\bibfield
  {journal} {\bibinfo  {journal} {Phys. Rev. D}\ }\textbf {\bibinfo {volume}
  {104}},\ \bibinfo {pages} {055042} (\bibinfo {year} {2021})},\ \Eprint
  {http://arxiv.org/abs/2104.14571} {arXiv:2104.14571 [hep-ph]} \BibitemShut
  {NoStop}%
\bibitem [{\citenamefont {Sirunyan}\ \emph {et~al.}(2019)\citenamefont
  {Sirunyan} \emph {et~al.}}]{CMS:2019ekd}%
  \BibitemOpen
  \bibfield  {author} {\bibinfo {author} {\bibfnamefont {A.~M.}\ \bibnamefont
  {Sirunyan}} \emph {et~al.} (\bibinfo {collaboration} {CMS}),\ }\href
  {\doibase 10.1103/PhysRevD.99.112003} {\bibfield  {journal} {\bibinfo
  {journal} {Phys. Rev. D}\ }\textbf {\bibinfo {volume} {99}},\ \bibinfo
  {pages} {112003} (\bibinfo {year} {2019})},\ \Eprint
  {http://arxiv.org/abs/1901.00174} {arXiv:1901.00174 [hep-ex]} \BibitemShut
  {NoStop}%
\bibitem [{\citenamefont {Langacker}(2009)}]{Langacker:2008yv}%
  \BibitemOpen
  \bibfield  {author} {\bibinfo {author} {\bibfnamefont {P.}~\bibnamefont
  {Langacker}},\ }\href {\doibase 10.1103/RevModPhys.81.1199} {\bibfield
  {journal} {\bibinfo  {journal} {Rev. Mod. Phys.}\ }\textbf {\bibinfo {volume}
  {81}},\ \bibinfo {pages} {1199} (\bibinfo {year} {2009})},\ \Eprint
  {http://arxiv.org/abs/0801.1345} {arXiv:0801.1345 [hep-ph]} \BibitemShut
  {NoStop}%
\end{thebibliography}
\end{document}